\documentclass[10pt]{JHEP3}
\pdfoutput=1
\usepackage{amsmath}
\usepackage{parskip}
\usepackage{verbatim}
\usepackage{graphicx}
\newcommand{\e}{\epsilon}

\renewcommand{\L}{{\mathcal{L}}}
\newcommand{\bL}{\bar{{\mathcal{L}}}}
\newcommand{\z}{{\bar z}}
\newcommand{\non}{\nonumber}

\newcommand{\be}[1]{ \begin{equation}\label{#1} }
\newcommand{\ee}{\end{equation}}
\newcommand{\bea}[1]{\begin{eqnarray}\label{#1} }
\newcommand{\eea}{\end{eqnarray}}
\newcommand{\bes}{\begin{subequations}}
\newcommand{\ees}{\end{subequations}}

\newcommand{\p}{\partial}

\newcommand{\refb}[1]{(\ref{#1})}

\newcommand{\<}{\langle}
\renewcommand{\>}{\rangle}
\renewcommand{\(}{\left(}
\renewcommand{\)}{\right)}
\renewcommand{\z}{{\bar z}}

\newcommand{\lb}{\left[}
\newcommand{\rb}{\right]}

\title{Galilean Conformal Electrodynamics}

\author{Arjun Bagchi$^{a,b}$, Rudranil Basu$^{a,c}$, Aditya Mehra$^a$
\\
$\;$ $\,$ $^a \, \,$ Indian Institute of Science Education and Research \\
$\;$ $\,$ $\,$ Dr Homi Bhabha Road, Pashan. Pune 411008. INDIA \\

$\;$ $\,$ $^b \, \,$ Institute of Theoretical Physics, Vienna University of Technology \\
$\;$ $\,$ $\,$ Wiedner Hauptstr. 8-10/136, A-1040 Vienna, Austria \\

$\;$ $\,$ $^c \, \,$ The Institute of Mathematical Sciences\\
$\;$ $\,$ $\,$Tharamani, Chennai 600113 INDIA \\

$\;$\email{a.bagchi@iiserpune.ac.in, rudrobose@gmail.com} \\
$\;$\email{aditya.mehra@students.iiserpune.ac.in}
}

\abstract{Maxwell's Electrodynamics admits two distinct Galilean limits called the Electric and Magnetic limits. We show that the equations of motion in both these limits are invariant under the Galilean Conformal Algebra in $D=4$, thereby exhibiting non-relativistic conformal symmetries. Remarkably, the symmetries are infinite dimensional and thus Galilean Electrodynamics give us the first example of an infinitely extended Galilean Conformal Field Theory in $D>2$. We examine details of the theory by looking at purely non-relativistic conformal methods and also use input from the limit of the relativistic theory.}

\preprint{}

\begin{document}

\baselineskip 3.5ex

\hfill
\newpage

\section{Introduction}
Throughout history, symmetries have been central to our attempts at understanding Nature and symmetry principles continue to play very important roles in the construction of modern theoretical physics. In recent times, conformal symmetry has been a principal actor in many diverse sub-fields. Conformal symmetry and field theories endowed with conformal symmetry, viz. conformal field theories (CFT) have been the key to understanding of apparently unrelated subjects like critical exponents in statistical mechanics, correlations in the cosmic microwave background and many different aspects of string theory and quantum gravity. 

Conformal symmetry is also known to appear as accidental symmetries of Maxwell's equations of electrodynamics in four spacetime dimensions. Although these symmetries become anomalous in the quantum regime, there are important lessons to be learnt here. When we look to generalise to Yang-Mills theories and then supersymmetrise to get to Supersymmetric Yang-Mills, conformal symmetries also survive in the quantum regime. 

One of the cornerstones of research in the past decade and half has been the AdS/CFT correspondence which has provided a concrete exemplification of the Holographic Principle in the context of string theory. The Holographic Principle \cite{'tHooft:1993gx, Susskind:1994vu} is the bold assertion that a theory of quantum gravity in a certain spacetime is completely equivalent to a theory without gravity living on the boundary of the same spacetime. The AdS/CFT correspondence realises this by relating Type IIB string theory on AdS$_5$ (times $S^5$) to $\mathcal{N} =4$ Super Yang-Mills (SYM) living on the boundary of AdS \cite{Maldacena:1997re}. The quantum conformal invariance of $\mathcal{N} =4$ SYM \cite{Beisert:2010jr} has been central to the enormous progress in the gauge-gravity duality. 

The true power of conformal symmetry manifests itself in two spacetime dimensions, where the symmetry algebra is enhanced to two copies of the infinite dimensional Virasoro algebra \cite{BPZ}. Armed with infinite dimensional symmetries, one can now compute various quantities of interest in conformally invariant systems without even resorting to a Lagrangian description. There has been interest in trying to implement general symmetry arguments to understand CFTs in D=3 and D=4. These include trying to understand classification of primary operators and conformally invariant operator product expansions, conformal blocks and the conformal bootstrap procedure. While some of this goes back a long way \cite{{Mack:1969rr, Ferrara:1973vz, Ferrara:1973yt}}, it has only recently been understood how to put these together to understand dynamical features about CFTs \cite{Rattazzi:2008pe, Rattazzi:2010yc, Poland:2011ey}. Of late there has been a resurgence of activities in this field with a special emphasis on trying to solve the 3D Ising model by general CFT methods like the conformal bootstrap \cite{ElShowk:2012ht}. 

In an ideal world, it would have been wonderful to have infinite symmetries in all spacetime dimensions which one could use to constrain various quantities of interest. This apparently utopian wish is actually fulfilled if one looks at Galilean Conformal Field Theories (GCFTs) which are the non-relativistic limits of conformal field theories \cite{Bagchi:2009my}. The relativistic conformal algebra undergoes an In{\"o}n{\"u}-Wigner contraction to the finite Galilean Conformal Algebra (f-GCA) with the same number of generators. It is then observed that the finite GCA can be given an infinite lift to an algebra which we will call the GCA (without the `f'). In two dimensions, the idea of an infinitely extended non-relativistic algebra is not unexpected and it can indeed be shown that the GCA is just a contraction of (linear combinations of) two copies of the Virasoro algebra \cite{Bagchi:2009pe}. But the infinite algebra in dimensions $D>2$ is indeed remarkable. 

To justify the claim that there is indeed an enhancement to infinite dimensional symmetries in non-relativistic conformal systems for $D>2$ and this is not just some mathematical jugglery, one needs physical examples. One partial example was provided in \cite{Bagchi:2009my} where it was mentioned that the non-relativistic Navier-Stokes equations of hydrodynamics realises a part of the symmetries of the GCA. But so far, there has been no concrete example of a GCFT in dimensions $D>2$. Our goal in this paper is to construct the first such example. For this we would be looking at the non-relativistic limit of Maxwell's electrodynamics in $D=4$. 

Galilean Electromagnetism was first discussed by Le Ballac and Levy-Leblond in 1973 \cite{LBLL}. They found that there were two distinct limits that they could perform on the theory, the Electric limit ($E_i >> B_i$) and the Magnetic limit ($B_i >> E_i$). Here, and elsewhere in this paper, we will set the speed of light $c=1$. In our discussions below, we shall reformulate this into a scaling of the spacetime co-ordinates and two different scalings of the components of the gauge field $A_\mu$. We shall show that both sets of equations of Galilean Electrodynamics (GED) in the Electric and Magnetic limits exhibit invariance under the f-GCA in $D=4$ and that there is a further infinite enhancement of symmetries to the full infinite dimensional GCA. 
We would be following recent work \cite{Jackiw:2011vz} which makes explicit the conformal structures of Maxwell's theory and would adapt our calculations to the non-relativistic case. Our paper would follow two parallel themes. We shall look at the limit of the appropriate analyses in the relativistic theory and also independently perform intrinsically non-relativistic calculations based on just the symmetries of the GCA.  The analysis just in terms of the symmetries of the non-relativistic theory is obviously more general, as we do not expect all GCFTs to descend as limits of CFTs, even in the 2D case \cite{Bagchi:2009pe}. We stress that the intrinsic approach is essential for $D>2$ in the studies of infinite extended GCA, which do not have any relativistic analogue. But, as we shall see, we need to resort to the limit to understand specifics of the particular example that we are dealing with. Certain constants which define the non-relativistic theory are determined through the limit from the relativistic theory.  

Before we go into the details of the computations, let us provide some further justification of why we are interested in this study over and beyond the obviously important goal of trying to find the first example of a non-relativistic conformal field theory in $D>2$ with infinite symmetries. As mentioned earlier, the quantum conformal invariance of $\mathcal{N} =4$ SYM is one of the main reasons why the AdS/CFT duality is so powerful. We expect that in the non-relativistic limit, Yang-Mills theory and subsequently Super Yang-Mills will continue to show invariance under the GCA. In particular, the hope is that in the supersymmetric setting, the non-relativistic limit of $\mathcal{N} =4$ SYM would exhibit GCA invariance also in the quantum domain. The existence of a non-relativistic sector with infinite dimensional symmetries raises very interesting possibilities. If this is a closed sub-sector in $\mathcal{N} =4$ SYM with infinite symmetries, it could very well turn out to be an integrable sub-sector and something that could be solved exactly, like the planar limit of $\mathcal{N} =4$ SYM. 

Integrability is usually a feature of two dimensions and the reason why the planar limit of $\mathcal{N} =4$ SYM is integrable is because this is linked through the AdS/CFT duality to the integrability of the string theory world-sheet which is a 2D non-linear sigma model. The non-relativistic subsector of $\mathcal{N} =4$ SYM could well be similarly linked to some limiting case of the string world sheet and hence exhibit similar features. It is interesting to point out that a tensionless limit of the closed string worldsheet actually exhibits symmetries of the 2D GCA \cite{Bagchi:2013bga}. It would be of great interest to explore these connections in more detail. But we will postpone discussions along these lines for future work. 

Let us now briefly outline the contents of this paper. In Sec 2, we revisit the analysis of \cite{Jackiw:2011vz, sheer} about the conformal invariance in relativistic electrodynamics which will help us set the stage for our non-relativistic ventures. In Sec 3, we recapitulate the construction of the GCA from the relativistic conformal symmetries. We then move on to constructing representations of the GCA. These are scale-spin representations as opposed to scale-boost ones which were considered earlier in \cite{Bagchi:2009ca} and play a central role in our analysis. In Sec 4, we first review the two distinct non-relativistic limits of \cite{LBLL} before reformulating this in the language of contractions of spacetime co-ordinates and gauge field components. The action of the conformal generators on the fields in the two limits is then studied. In Sec 5, we demonstrate the invariance of the two limiting theories under first the finite part of the GCA and then exhibit the enhancement to the full infinite dimensional symmetry. We go on to construct correlation functions for the gauge fields in both sectors. In Sec 6, we turn our attention to the formulation of electrodynamics in 3D with a scalar field and show that its non-relativistic limit again exhibits GCA invariance which is again infinite dimensional. This provides us with another example of a GCFT, now in $D=3$. We close with a summary of results, connections to other related work and more future directions in Sec 7. 


\section{Conformal Invariance and Electrodynamics}

Conformal invariance, as stressed in the introduction, is of great importance in many different branches of theoretical physics. In general $D$ space-time dimensions, the conformal group consists of the Poincare group along with scaling and special conformal transformations (SCT). There has been a debate in the literature on whether the Poincare group along with scaling necessarily leads to conformal invariance. While this hold true in dimensions $D=2$ for well-behaved quantum field theories (unitary, with a well defined energy-momentum tensor and a discrete spectrum) \cite{Zamolodchikov:1986gt, Polchinski:1987dy}, there are examples of scale invariant, non-conformal systems in higher dimensions, with classical electrodynamics being one of the simplest examples \cite{Jackiw:2011vz, sheer}. 

Before we go into the details of conformal invariance in Maxwellian Electrodynamics, in order to set some notation, let us remind the reader of some details of the conformal algebra in $D=4$. The Poincare sub-algebra consists of spacetime translations ($\tilde P_i= \p_i, \tilde H=-\p_t$), rotations ($\tilde J_{i j} = x_i \p_j - x_i \p_j$) and boosts ($ \tilde B_i = x_i \p_t + t \p_i$). The other generators are the Dilatation operator ($\tilde D = -x \cdot \p$) and the Special conformal generators ($\tilde K_\mu =  - 2 x_\mu (x \cdot \p)  + (x \cdot x) \p_\mu$). The conformal algebra in $D$ dimensions is isomorphic to $\mathfrak{so}(D,2)$. Let us indicate a few important commutation relations below which will help us highlight the differences between the relativistic case considered here and the non-relativistic case that would be the central theme of this paper:
\be{rel-al}
[\tilde P_i, \tilde B_j]= - \delta_{ij} \tilde H, \, \, [\tilde B_i, \tilde B_j] =  \tilde J_{ij}, \, \, [\tilde K_i, \tilde B_j] = \delta_{ij} \tilde K, \, \, [\tilde K_i, \tilde P_j] = 2 \tilde J_{ij} + 2 \delta_{ij} \tilde D. 
\ee
We now go on to describe, following \cite{Jackiw:2011vz}, the transformation laws of fields under conformal transformations. Let us consider transformations of a multi-component field $\Phi$. For Poincare transformations, which leave the action for $\Phi$ invariant,
we have
\be{}
{\tilde{\delta}}_\mu^{\mbox{\tiny{trans}}} \, \Phi (x) = \p_\mu \Phi (x), \quad {\tilde{\delta}}_{\mu \nu}^{\mbox{\tiny{Lorentz}}} \, \Phi (x) = (x_\mu \p_\nu - x_\nu \p_\mu + \Sigma_{\mu\nu}) \Phi(x)
\ee
Transformations under scaling takes the following form:
\be{}
{\tilde{\delta}}^{\mbox{\tiny{scale}}}  \, \Phi (x) = (x \cdot \p + {\tilde{\Delta}}) \Phi
\ee
where ${\tilde{\Delta}}$ is the scaling dimension of the field $\Phi$ and to make the kinetic term of the corresponding action scale invariant one chooses
\be{}
{\tilde{\Delta}} = \frac{D-2}{2}.
\ee
The transformation under special conformal transformation can be obtained by looking the commutation relations of the conformal algebra. Specialising to the case of primary fields, we find that 
\be{}
{\tilde{\delta}}^{\mbox{\tiny{SCT}}}_\mu  \, \Phi (x) = \{ 2 x_\mu (x \cdot \p) - x^2 \p_\mu + 2 {\tilde{\Delta}} x_\mu - 2 x^\nu \Sigma_{\nu \mu} \}  \, \Phi (x).
\ee
We shall also be considering non-primary or descendent fields, like derivatives of primary fields, in our analysis. The transformation laws of these fields can be obtained from the transformations of their corresponding primary fields. 

Let us now consider free Maxwell theory in D-dimensional spacetime. The theory is best expressed in terms of a field strength $F_{\mu\nu} = \p_\mu A_\nu - \p_\nu A_\mu$ where $A_\mu$ is the fundamental dynamic variable, the gauge field.  $F_{\mu \nu}$ satisfies
\bea{eombi}
\mbox{{Equations of Motion:}} \qquad && \p^\mu F_{\mu \nu} = 0 \\
\mbox{{Bianchi identity:}} \qquad && \p_\rho F_{\mu \nu} + \p_\nu F_{\rho \mu} + \p_\mu F_{\nu \rho} = 0.  
\eea
The equations of motion can be derived from the well-known Lagrangian
\be{L}
\mathcal{L} = - \frac{1}{4} F_{\mu \nu} F^{\mu \nu}
\ee
While the above Lagrangian is manifestly invariant under Poincare transformations:
\bes \label{poinc}
\bea{}
&&{\tilde{\delta}}_\mu^{\mbox{\tiny{trans}}} \, A_{\nu} (x) = \p_\mu A_{\nu} (x) \\ &&{\tilde{\delta}}_{\mu \nu}^{\mbox{\tiny{Lorentz}}} \, A_{\rho} (x) = (x_\mu \p_\nu - x_\nu \p_\mu)A_{\rho}(x) + g_{\rho \mu} A_{\nu}(x)-g_{\rho \nu} A_{\mu}(x),
\eea
\ees
scale and special conformal transformations act non-trivially on it. In terms of the field variable $A_\mu$, the transformations are the following
\bea{transA}
{\tilde{\delta}}^{\mbox{\tiny{scale}}}  \, A_\mu (x) &=& (x \cdot \p + {\tilde{\Delta}}) A_\mu (x) \\
{\tilde{\delta}}^{\mbox{\tiny{SCT}}}_\mu  \, A_\nu (x) &=& \{2 x_\mu (x \cdot \p) - x^2 \p_\mu + 2 {\tilde{\Delta}} x_\mu \} A_\nu (x) - 2 x_\nu A_\mu (x) +  g_{\mu \nu} x^\rho A_\rho (x),
\eea
where the conformal dimension of the field is ${\tilde{\Delta}} = \frac{D-2}{2}$. The last line is the conformal transformation where the gauge field has been specialised to a primary field. While to prove conformal invariance, it is important to show that the Lagrangian \refb{L} is invariant up to a total divergence under conformal transformations, we shall concentrate on understanding the conformal invariance of the equations of motion \refb{eombi}. This is because in the non-relativistic setting in which we are interested, there is no straightforward action or Lagrangian that one can construct. We would thus focus on the symmetries of the equations of motion. For this we require the transformations of the field strength. These can be derived directly from the transformations of the gauge fields that we have outlined above in \refb{transA}. Under scaling and special conformal transformations, the field strength transforms as: 
\bea{}
&&{\tilde{\delta}}^{\mbox{\tiny{scale}}}  \, F_{\mu \nu} (x) = (x \cdot \p + \frac{D}{2}) F_{\mu\nu} (x) \\
&&{\tilde{\delta}}^{\mbox{\tiny{SCT}}}_\sigma  \, F_{\mu \nu} (x) = \Xi_\sigma F_{\mu \nu} + (D-4) (g_{\sigma \alpha} A_\beta - g_{\sigma \beta } A_\alpha) \\
&& \mbox{where} \quad \Xi_\sigma F_{\mu \nu} = (2 x_\sigma x_\tau - g_{\sigma \tau} x^2) \p^\tau F_{\mu \nu} + D x_\sigma F_{\mu \nu} + 2 (g_{\sigma \mu} x^\tau F_{\tau \nu} + g_{\sigma \nu} x^\tau F_{\mu \tau} -  x_\mu F_{\sigma \nu} -  x_\nu F_{\mu \sigma}). \nonumber
\eea
Following \cite{Jackiw:2011vz}, one can show that the Lagrangian \refb{L} is invariant (up to total derivative terms) under dilatations. In the case of special conformal transformations, this is true only in dimension $D=4$. The equations of motion show a similar behaviour. If they are to be invariant under special conformal transformations, $\p^\mu \, {\tilde{\delta}}^{\mbox{\tiny{SCT}}}_\sigma  \, F_{\mu \nu}$ should be zero. It can be shown by considering the above transformations that 
\be{}
\p^\mu \, {\tilde{\delta}}^{\mbox{\tiny{SCT}}}_\sigma  \, F_{\mu \nu} = (D-4) (\p_\nu A_\sigma - g_{\nu \sigma} \p^{\tau} A_{\tau}) \neq 0 \quad {\mbox{for}} \quad D\neq 4
\ee
This implies that Maxwell's equations are only conformally invariant in $D=4$. In dimensions $D>4$, Maxwellian electrodynamics forms an example of a scale invariant but non-conformal relativistic theory. In $D=3$, a special formulation of electrodynamics is possible where one can formulate everything in terms of a scalar potential and recover conformal symmetry \cite{Jackiw:2011vz}. We shall return to this discussion near the end of this paper. 

Conformal invariance of Maxwell's equations in $D=4$ or the lack of it in $D>4$ can be expressed in terms of correlation functions of the theory starting from the Lagrangian \refb{L} \cite{sheer}. Although, we shall not be using this approach in our non-relativistic approach, it is good to keep this in mind.  We eventually hope to have a Lagrangian description which would possibly depend on Newton-Cartan structures as outlined briefly near the end of the paper and this approach should prove valuable there. In position space in $D=4$, the two-point function of gauge fields is given by 
\be{em-corr}
\< A_\mu (x) \, A_\nu(0) \> = \frac{g_{\mu\nu}}{x^2} + \mbox{gauge dependent terms} 
\ee
Since this is gauge dependent, the correlator is obviously not a physical one, but it helps to define the two-point function of the gauge independent field strengths which in turn defines the theory.
\be{F2pt}
\< F_{\mu \nu} F_{\rho\sigma} \> = \frac{I_{\mu \rho} I_{\nu\sigma} - I_{\mu \sigma} I_{\nu \rho}}{x^4}, \quad \mbox{where} \, \, I_{\mu\nu} = g_{\mu \nu} - \frac{2 x_\mu x_\nu}{x^2}
\ee
The stress energy tensor can be constructed out of the field strength: 
\be{Tem}
T_{\mu \nu} = F_{\mu \rho} F^{\rho}_{\, \nu} - \frac{1}{4} g_{\mu \nu} F_{\rho \sigma} F^{\rho \sigma}.
\ee
The common way of arriving at the above expression is by varying the Lagrangian \refb{L} with respect to the metric. But \cite{sheer} argues that it is enough to look at the expression \refb{Tem} and construct its correlation functions and check if they are conserved and satisfy Ward identities. In order to now check whether the Maxwellian theory is conformally invariant, one can compute the trace of the energy momentum tensor which in $D=4$ turns out to be zero. This however is not as straight-forward in dimensions other that four. Although one can straightaway check that the trace of the EM tensor defined through \refb{Tem} is non-zero when $D\neq 4$:
\be{}
T^\mu_{\, \mu} = \frac{D-4}{4} F_{\mu \nu} F^{\mu \nu} 
\ee
one can add ``improvement" terms to the EM tensor which may potentially make the trace zero. It can be argued that this is not possible for Maxwell's theory \cite{sheer}. Another way of testing the presence of conformal symmetry is to look at the correlators of the theory. One can look at the operator $\Phi = \, :(F_{\mu \nu})^2: $ and look for example at its three point function which in a conformal field theory is fixed by symmetries. The three point function $\< \Phi(x) \Phi(y) \Phi(z) \>$ actually vanishes in $D=4$ and this is consistent with the assertion that this is a primary in the CFT. The structure of 
\refb{F2pt} is also consistent with the two point function of a tensor valued anti-symmetric primary.  

We shall not be adopting this approach as our non-relativistic construction does not have a straight forward Lagrangian description and hence the construction of correlation functions is a problem. In the non-relativistic conformal case in $D=4$, we would be able to construct correlation functions from the symmetries. But presently, we do not have a way of cross-checking the forms of the correlators with explicit answers from an action principle. 


\section{Non-relativistic Conformal Symmetries}

\subsection{Contraction and Infinite Extension}

We wish to consider the non-relativistic equivalent of the conformal symmetry. For this we shall consider the Inonu-Wigner contraction of the conformal algebra. We shall be performing this limit at the level of the spacetime and the contraction will be performed in units where the speed of light $c=1$. In the non-relativistic limit, velocities would be small compared to the speed of light and hence in these units, it should correspond to $v_i \to 0$. We shall achieve this by the parametric contraction 
\be{contr}
x_i \to \e x_i, \, \, t \to t, \,\, \e \to 0 
\ee
so that $v_i \sim \frac{x_i}{t} \to 0$. In a contraction, the number of generators does not change from the parent to the daughter algebra, but the expressions for some of the generators change. Let us demonstrate this for one example below.
\be{}
\tilde B_i = t \p_i + x_i \p_t \to \frac{1}{\e} t \p_i + \e x_i \p_t \Rightarrow  B_i = \lim_{\e \to 0} \e \tilde B_i = t \p_i 
\ee 
The algebra we will get after the contraction will be called the finite Galilean Conformal Algebra (f-GCA). The generators of the f-GCA are 
\bea{fgcagen}
&& H = -\p_t, \quad D = -(t \p_t + x_i \p_i), \quad K = -(t^2 \p_t + 2 x^i t \p_i) \\
&& P_i = \p_i, \quad B_i = t \p_i, \quad K_i = t^2 \p_i \\
&& J_{ij} = x_{[i} \p_{j]}
\eea
where $[]$ is the anti-symmetrizer and is used without any combinatorial factor.

Here the Galilean subalgebra is generated by $\{H, P_i, B_i, J_{ij} \}$. When we compute the commutators, we see that the difference with the relativistic conformal algebra arises in the analogues of \refb{rel-al}. The right-hand sides of all these commutators vanish in the non-relativistic limit. 
We see that the generators can be written in the suggestive form for $n=0, \pm 1$: 
\bea{suggen}
&& L^{(n)} = -t^{n+1} \p_t - (n+1) t^n x_i \p_i \quad ( L^{(-1, 0, +1)} = H, D, K)\\
&& M^{(n)}_i = t^{n+1} \p_i \quad ( M^{(-1, 0, +1)}_i = P_i, B_i, K_i)
\eea
The identifications with the contracted generators \refb{fgcagen} are written in parenthesis. The f-GCA can be written as
\bea{GCA}
&& [L^{(n)}, L^{(m)}] = (n-m) L^{(n+m)}, \quad [L^{(n)}, M^{(m)}_i] = (n-m) M_i^{(n+m)} \quad [M^{(n)}_i, M^{(m)}_j]=0 \\
&& [L^{(n)}, J_{ij}] = 0, \quad [M^{(n)}_i, J_{jk}] = M^{(n)}_{[k} \delta^{}_{j]i}
\eea
We now observe that the above algebra continues to hold even if we consider generators \refb{suggen} for any integer value of $n$. 
{\footnote{In earlier work, it has been noted that even the rotation generators could be given an infinite lift following
\be{}
J_a^{(n)} \equiv J_{ij}^{(n)}  = t^{n} (x_i\p_j - x_j \p_i)
\ee  
These generators constitute the following infinite dimensional algebra
\bea{}
&& [L^{(n)}, L^{(m)}] = (n-m)  \, L^{(n+m)}, \quad [L^{(n)}, M^{(m)}_i] = (n-m) \, \, M_i^{(n+m)}, \quad [M^{(n)}_i, M^{(m)}_j]=0 \non \\
&& [J_a^{(n)}, J_b^{(m)}] = f_{ab}{}^{c} \, \, J_c^{(n+m)}, \quad [L^{(n)}, J_a^{(m)}] = m \, \, J_a^{(n+m)} \quad  [M^{(n)}_i, J^{(m)}_{jk}] = M_{[k}^{(n+m)} \delta^{}_{j]i}.
\eea
In this paper, we would not be considering the infinite dimensional lift for the rotation generators. This is primarily because of some complications in the analysis to follow which we would like to avoid for the present considerations. It is important to note that in the supersymmetric case \cite{Bagchi:2009ke}, it has been noted that the rotations cannot be given this infinite lift. Since our ultimate goal is to look at Supersymmetric Yang-Mills theories, the present restriction would not be an issue when one generalises to the supersymmetric analysis. }}
The generators $\{ L_n \}$ form a sub-algebra which with suitable central extensions becomes a Virasoro algebra. The symmetry algebra above \refb{GCA} is an infinite dimensional algebra for any space-time dimension. We have hence the tantalising prospect of infinite symmetry enhancements in the non-relativistic limit of a conformal field theory in any dimensions. 

The obvious question is whether this infinite dimensional symmetry is realised in any physical example. As we will go on to describe next, in $D=2$, there are surprising appearances of this algebra in different apparently unconnected fields. In $D>2$, we have the example of hydrodynamics. The non-relativistic Navier-Stokes equation 
is invariant under arbitrary time-dependent accelerations, the $M_i^{(n)}$'s of the above algebra, thus realising an infinite dimensional sub-set. The viscous term in the Navier-Stokes introduces a scale in the system and hence the dilatation operator is no longer a symmetry. So the Navier-Stokes equation only realises a part of the infinte dimensional GCA. If we turn off viscosity to move to the Euler equation, the f-GCA becomes a symmetry. But the higher $L_n$ modes do not leave the equation invariant.

So, before the present paper, there have been no examples in $D>2$ of systems realising the full infinite dimensional GCA as their symmetry algebra. 

\subsection{The case of $D=2$}\label{d2}

In $D=2$, the appearance of an infinite non-relativistic conformal algebra is much less surprising given the fact that the 2d relativistic conformal algebra itself is enhanced to two copies of the infinite dimensional Virasoro algebra. The 2d GCA can actually be obtained by contraction of linear combinations of these two Virasoro algebras. If we denote $\L_n, \bL_n$ as the two copies of relativistic Virasoro:
\bea{gca2}
&& [\L_n, \L_m] = (n-m) \L_{n+m} + \frac{c}{12} \delta_{n+m,0} (n^3 - n), \quad [\L_n, \bL_m] = 0 \\
&& [\bL_n, \bL_m] = (n-m) \bL_{n+m} + \frac{\bar{c}}{12} \delta_{n+m,0} (n^3 - n)
\eea
then the 2d GCA is obtained by looking at the limit 
\be{NR}
L_n = \L_n + \bL_n, \quad M_n = \e ( \L_n - \bL_n).
\ee
This leads to the 2d GCA:
\bea{2dGCA}
 && [L_n, L_m] = (n-m) L_{n+m} + \frac{c_L}{12} \, \delta_{n+m,0} (n^3 - n), \\ 
 && [L_n, M_m] = (n-m) M_{n+m} + \frac{c_M}{12} \, \delta_{n+m,0} (n^3 - n), \quad [M_n, M_m]=0 \nonumber
\eea
It is interesting to note that in this case, the GCA in 2d admits a central extension between $L_n$ and $M_m$. This in higher dimensions would be a vector and hence not a central charge. The two central extensions are related to their relativistic versions by the relations: 
\be{}
c_L = c + \bar{c}, \quad c_M = \e ( c - \bar{c} ) \nonumber.
\ee
The fact that \refb{NR} is a non-relativistic limit can be understood by looking at the familiar representations of the Witt algebra in terms of holomorphic and anti-holomorphic co-ordinates, rewriting them in terms of $x, t$ and performing the contraction  \refb{contr}
\be{plane}
\L_n = z^{n+2} \p_z, \quad \bL_n = \z^{\, n+2} \p_\z; \quad z, \z = t \pm \e x 
\ee
The 2d GCA can also be obtained by another contraction which combines the generators as 
\be{UR}
L_n = \L_n - \bL_{-n}, \quad M_n = \e ( \L_n + \bL_{-n}).
\ee
It can be easily checked that again the commutators \refb{2dGCA} are reproduced. The identifications of the central terms also change to $c_L = c - \bar{c}$ and $c_M = \e (c+ \bar{c})$. From the point of view of spacetime, \refb{UR} actually can be motivated from an ultra-relativistic contraction as opposed to the non-relativistic one \refb{NR}. For this, it is important to look at the equivalent of the expressions \refb{plane} on the cylinder and look at the contraction which sends the speed of light to zero instead of infinity. This contraction is important for considerations in flat holography. It turns out that the asymptotic symmetry algebra of flat space in three dimensions \cite{Bondi:1962px, Sachs:1962zza, Barnich:2006av} is isomorphic to the 2d GCA and hence a dual field theory to 3d flat spacetimes is actually a Galilean Conformal Field Theory. This fact has been exploited to build aspects of flat holography in 3d \cite{Bagchi:2010eg}. A incomplete set of references include \cite{Bagchi:2010eg} -- \cite{Duval:2014lpa}. It is interesting to point out here that there has been the construction of an explicit field theory example where the flat limit of Liouville theory was performed and the Poisson algebra of the conserved charges realise the infinite dimensional algebra \refb{2dGCA} \cite{Barnich:2012rz}. 

Apart from the uses in non-relativistic physics and also in the development of flat space holography in 3D, it has also been noticed that the 2D GCFTs would have a central role to play in the tensionless limit of bosonic string theory. The symmetries \refb{2dGCA} arise as residual symmetries on the world sheet in the ``conformal" gauge of the tensionless closed string, and this is analogous to the appearance of the 2d conformal symmetry in the case of the usual tensile bosonic closed string. 

The unexpected appearance of the GCA in various apparently unrelated fields of physics seem to point to a universality in the behaviour of limits of conformally invariant systems. It is increasingly becoming of importance to explore these symmetries and our current paper is another very important step in this direction as it deals with the appearance of a GCFT in its full infinite dimensional form, now in a higher dimensional example.  

\subsection{Representations of GCA: Scale-Spin Primaries}\label{repsec}

The representations of the GCA can be constructed along lines similar to that of the relativistic conformal algebra.  We will label the states with their weights under the dilatation operator $L^{(0)}$. 
\be{scale}
L^{(0)} | \Phi \> = \Delta | \Phi \>
\ee
We notice that the weights are lowered by the operators $L^{(n)}, M_i^{(n)}$ for positive $n$ and raised by ones with negative $n$. 
\be{}
L^{(0)} L^{(n)} | \Phi \> = (\Delta -n) L^{(n)} | \Phi \>, \quad L^{(0)} M_i^{(n)} | \Phi \> = (\Delta -n) M_i^{(n)} | \Phi \>.\nonumber
\ee
This helps us build primary states in the theory in analogy with CFTs by demanding that the spectrum be bounded from below. We define primary states as 
\be{posmode}
L^{(n)} | \Phi \>_p = M_i^{(n)} | \Phi \>_p = 0 \quad \forall n >0
\ee
We now have two different options in labelling these states further. Since 
\be{}
[L^{(0)}, M_i^{(0)}] = 0 \quad [L^{(0)}, J_{ij}] = 0 \quad [M_i^{(0)}, J_{ij}] \neq 0, \nonumber
\ee
we can choose to label the states further by the weights under $M_i^{(0)}$ or $J_{ij}$, but not with both of them. In \cite{Bagchi:2009ca}, we chose to label the primaries with $M_i^{(0)}$ and this was also useful in 2D \cite{Bagchi:2009my} as this was identified with $\L_0 - \bL_0$ in the non-relativistic limit. But in this paper, we shall be interested in understanding operators with spin \footnote{At this stage, by `spin' we would mean the weight vector or the set of all quantum numbers specifying a particular irreducible representation of $\mathfrak{so}(D-1)$ generated by $J_{ij}$ for $D \geq 4$} and hence we will look at primaries which are labelled under $J_{ij}$. 
\be{}
J_{ij}^{(0)} | \Phi \>_p = \Sigma_{ij} \rhd | \Phi \>_p
\ee
where $\Sigma_{ij} \rhd$ denotes action of $J_{ij}$ in that particular representation of $SO(D-1)$, to which $ |\Phi \>_p$ belongs. The representations of the GCFT are built by acting with the raising operators ($L^{(n)}, M_i^{(n)}$ for negative $n$) on these primary states. 

For sake of explicitly studying the action of GCA on the operators of our theory, we would use the state-operator correspondence and would denote by $ \Phi (t,x)$ primary operator(s) relating the primary state and the vacuum:
\be{}
| \Phi \>_p = \Phi (0,0) | 0 \>.
\nonumber
\ee
We would assume as usual, the action of the finite part of GCA on these primaries consistently with \eqref{posmode}, \eqref{scale} as:
\bes 
\bea{}
&&\left[L^{(-1)}, \Phi(t,x)\right] = \partial _t \Phi(t,x) \label{ham} \\
&&\left[M^{(-1)}_i, \Phi(t,x)\right]= -\partial _i \Phi(t,x) \label{mom}\\
&&\left[J_{ij}, \Phi(0,0)\right] =  \Sigma _{ij} \rhd \Phi(0,0) \\
&&\left[L^{(0)}, \Phi(0,0)\right] = \Delta \Phi(0,0) \\
&&\left[L^{(+1)}, \Phi(0,0)\right] = 0 = \left[M^{(+1)}_i, \Phi(0,0)\right]
\eea
\ees
Let's now briefly recapitulate how the actions of $J_{ij}, L^{(0,1)}$ and $M^{(+1)}_i$ can be worked out on an operator at general space-time points $(t,x)$. According to our conventions,
\be{}
\Phi(t,x) = U \Phi(0,0) U^{-1}
\ee
where $U= e^{(t L^{(-1)} - x^i  M^{(-1)}_i)}$. For a general GCA element $ \bigstar$, we therefore use:
\be{}
\left[ \bigstar, \Phi(t,x)\right] = U \left[ U^{-1} \bigstar U , \Phi(0,0)\right] U^{-1} \nonumber
\ee
and then exploit the Baker-Campbell-Hausdorff formula (BCH) and GCA \eqref{GCA} to evaluate $U^{-1} \bigstar \, U$. This is straightforward for $J_{ij}$ and $L^{(0)}$:
\bea{JD}
&& \left[J_{ij}, \Phi(t,x)\right] =  x_{[i} \p_{j]} \Phi(t,x)+ \Sigma _{ij} \rhd \Phi(t,x) \non \\
&& \left[L^{(0)}, \Phi(t,x)\right]= (t \partial _t +x^i \partial _i + \Delta) \Phi (t,x)
\eea
However we encounter a problem as soon as we try to find the action of $L^{(+1)}$ and $M^{(+1)}_i$, i.e. the generators of Galilean version of special conformal transformation. This is because we need the role of boost operator for these evaluations. We have avoided it up to this point, since we do not want boost to label our primaries. This is the opportune moment to digress on the action of $B_i$ (which is $M^{(0)}_i$ according to our notation).

In this direction, let us look at the following equation found from the relevant Jacobi identity:
\be{} \label{jacspin}
\left[J_{ij}, \left[ B_k , \Phi (0,0)\right] \right]=  \lb B_k, \Sigma_{ij} \rhd \Phi (0,0) \rb + \delta _{k [i} \lb B_{j]}, \Phi (0) \rb .
\ee
Moreover, a similar Jacobi identity guarantees that $B_i$ preserves scaling dimension of $\Phi$. With Galilean ED as a motivation, we study following two cases explicitly.
\begin{itemize}

\item \textit{\underline{ $\Phi$ is $\mathfrak{so}(D-1)$ scalar}}

Let $ \phi$ denote $\Phi (0,0)$ here. Hence from \eqref{jacspin} we infer:
\be{spinscal}
\left[J_{ij}, \left[ B_k , \phi \right] \right]= \delta _{k[i} \lb B_{j]}, \phi \rb
\ee 
In its most general form:
\be{}
\lb B_i, \phi \rb = a_i \,\phi + b \,\phi _i + \mbox{ higher spin operators} 
\ee 
where $a_i$ is a c-number valued vector and $ \phi _i$ is a vector operator (spin-1 for $D=4$) at space-time origin. Note that presence of $a_i$ as a c-number valued vector breaks the $SO(D-1)$. Hence, we must have $a_i=0$ and
\be{bphi}
\lb B_i, \phi \rb =  b\, \phi _i + \mbox{ higher spin operators}
\ee 
Structure of the higher spin (tensor) operators in general could only have a term of the form: $ \varepsilon_i{}^{i_1 i_2 \hdots i_{D-2}}\, \phi _{i_1 i_2 \hdots i_{D-2}}$, since $\varepsilon$ is the only $\mathfrak{so}(D-1)$ invariant tensor other than $ \delta_{ij}$. However, the right hand side of \eqref{jacspin} transforms in a vector representation, negating the possibility of any higher spin operator as a consequence. On the other hand, if one prefers not to have rotation as an invariance, $a_i$ essentially measures the `rapidity' of $ \phi$, in case $ b=0$, as in \cite{Bagchi:2009ca}.

One must keep note of the fact that both the scalar $\phi$ and the vector it transforms into $ \phi _i$, share the same scaling dimension.

\item \textit{\underline{ $\Phi$ is $\mathfrak{so}(D-1)$ vector}}

Let $ \tilde{\phi}_i$ denote $\Phi (0,0)$ here. Hence from \eqref{jacspin} we have:
\be{case2} 
\left[J_{ij}, \left[ B_k , \tilde{\phi}_l \right] \right]= \lb B_k , \delta_{l [ i} \tilde{\phi}_{j]} \rb + \delta _{k[i} \lb B_{j]}, \tilde{\phi}_l \rb
\ee
We use the rotational symmetry arguments presented in the earlier case again to solve this most generically by 
\be{boosta}
\lb B_i, \tilde{\phi} _j \rb = a \delta_{ij} \tilde{\phi},
\ee 
where $a$ is a c-number and $\tilde{\phi}$ is a spin-less operator. In general these $\tilde{\phi}$ and $ \tilde{\phi}_i$ need not have to do anything with the $\phi$ and $ \phi_i$ appearing in the previous case. However, if we demand (which we should, for the case of concrete example of Galilean Electrodynamics) $ \phi$ and $\phi_i$ to be of same scaling dimension, then $ \phi \sim \tilde{\phi}$ and $ \phi_i \sim \tilde{\phi}_i$ upto some real multiplier, which we choose to be 1 in what follows.
\end{itemize}
We see that we are yet to determine the constants $a,b$ appearing in \eqref{bphi}, \eqref{boosta} using all of kinematical symmetry. Another way of saying this is that our representations are labelled by two constants, in addition to weight and spin. It will become apparent in the following analysis that $a,b$ get fixed for particular physical systems. Determination of these constants demand inputs from dynamics.

With the above information at hand, it is easy to generalize the action of boost on operators at finite space-time points:
\bes{}\label{boostfin}
\bea{}
\lb B_i, \phi (t,x) \rb &=& -t \partial_i \phi (t,x) + b \phi _i (t,x) \\
\lb B_i, \phi_j (t,x) \rb &=& -t \partial_i \phi_j (t,x) + a \delta_{ij} \phi (t,x) 
\eea
\ees
Now we are ready to study the action of $L^{(+1)}$ and $ M_i^{(+1)}$ on our scalar and vector primaries of weight $ \Delta$ at arbitrary space-time points. We use the scheme of using BCH and \eqref{GCA} outlined above in deriving \eqref{JD},\eqref{boostfin}:
\bes{}\label{sctgen}
\bea{} 
\label{dummy}\lb L^{(+1)}, \phi(t,x) \rb &=& (t^2 \partial _t + 2 t x^i \partial_i + 2t \Delta) \phi (t,x)  - 2b\, x^i \phi_i (t,x) \\
\lb L^{(+1)}, \phi_i(t,x) \rb &=& (t^2 \partial _t + 2 t x^i \partial_i + 2t \Delta) \phi_i (t,x) - 2a \, x^i \phi (t,x) \\
\lb M_i^{(+1)}, \phi(t,x) \rb &=& - t^2  \partial _i  \phi (t,x) + 2 b\, t  \phi _i (t,x)\\
\lb M_i^{(+1)}, \phi _j(t,x) \rb &=& - t^2  \partial _i  \phi_j (t,x) + 2 a\,t \delta_{ij} \phi (t,x)
\eea
\ees
Let's pause for a moment to take stock of the progress made so far in accomplishing our goal of studying action of GCA on scale-spin primaries. From the equations \eqref{ham}, \eqref{mom}, \eqref{JD}, \eqref{boostfin} and \eqref{sctgen} we conclude that for the global or finite part of GCA, we have successfully produced what we desired to, additionally labelled by two numbers $a,b$. Conceptually, it should not be hard to do the same for infinite extension of GCA \eqref{GCA}. The essentially non-trivial computational step lies in simplification of adjoint action of Galilean Hamiltonian and momentum operators on arbitrary modes of GCA:
$$ U^{-1} \bigstar ^{(n)} U.$$
This can be performed again appealing to BCH and keeping in mind that $U= e^{(t L^{(-1)} - x^i  M^{(-1)}_i)}$ and repeatedly using \eqref{GCA} \cite{Bagchi:2009ca}. To be more concrete, for non-negative modes:
\bes{}
\bea{adj}
U^{-1} L ^{(n)} U &=& \sum _{k=0}^{n+1} {{n} \choose {k}} \( t^k \, L^{(n-k)} - k \, t^{k-1} x^i M_i^{(n-k)}\) \\
U^{-1} M^{(n)}_i  U &=& \sum _{k=0}^{n+1} {{n} \choose {k}} t^k \, M^{(n-k)}_i 
\eea
\ees
Using these, we arrive at (here we have assumed, as usual, that the positive modes of GCA generators annihilate the primaries.):
\bes\label{inf}
\bea{}
\lb L^{(n)}, \phi(t,x) \rb &=& t^n \(t \p _t + (n+1) x^i\, \p_i + (n+1)\Delta \) \phi(t,x) \non \\ &-& b\,n\,(n+1)t^{n-1}\,x^i \phi _i (t,x) \\
\lb L^{(n)}, \phi_i(t,x) \rb &=& t^n \(t \p _t + (n+1) x^j\, \p_j + (n+1)\Delta \)\phi_i(t,x)\non \\ &-& a\,n\,(n+1)t^{n-1}\,x_i \phi(t,x) \\
\lb M^{(n)}_i, \phi(t,x) \rb &=& -t^{n+1} \p_i \phi (t,x)+ b\,(n+1)\,t^{n} \phi_i(t,x) \\
\lb M^{(n)}_i, \phi_j(t,x) \rb &=& -t^{n+1} \p_i \phi_j (t,x)+ a\,(n+1)\,t^{n} \delta_{ij}\phi(t,x)
\eea
\ees
With these at hand, we have successfully constructed modules defined by the quadruplet $ \Delta, \Sigma, a, b$ serving as representations of GCA. 


\section{Galilean Electrodynamics}
In this section, following the work by Le Bellac and Levy-Leblond \cite{LBLL}, we first briefly review the Galilean version of Maxwell's electrodynamics. Then we reformulate the same in terms of scalings of the spacetime and the components of the gauge field. We then go on to describe the action of the Galilean Conformal generators on GED by implementing this limiting procedure from relativistic electrodynamics. 

\subsection{The Tale of Two Limits}
The fundamental observation of \cite{LBLL} was that Maxwell's theory admits {\em{two different non-relativistic limits}}. This can be understood from the usual Lorentz transformation of a four-vector $u_\mu = (u_0, u_i)$ which admits two different Galilean limits:
\bea{}
u_0' = u_0, \quad u_i' = u_i - v_i  u_0 \label{Etype} \\
u_0' = u_0 - v^i u_i, \quad  u_i' = u_i \label{Btype}
\eea
Here we have set the speed of light $c=1$ and $v_i$ is the velocity of the transformation. \refb{Etype} applies when $v<<1$ and $u_0 >> \sqrt{u_i^2}$ and \refb{Btype} applies when $v<<1$ and $u_0 << \sqrt{u_i^2}$. In the case of Electromagnetism, \refb{Etype} would correspond to the case where $|\rho | >> |\vec{j}|$ and is called the Electric limit. In this limit, true to its name, the electric effects dominate over the magnetic ones. \refb{Btype} applies when $|\rho | << |\vec{j}|$ and is called the Magnetic limit. Here magnetic effects dominate over the electric ones. It is straight-forward to write down the analogues of the Maxwell equations in these limits. These are displayed below.
\bea{}
&&\mbox{\underline{Electric limit:}} \nonumber \\
&& \p_i E^e_i = \frac{\rho_e}{\varepsilon_0}, \quad \e_{ijk}\p_j E^e_k = 0, \quad \p_i B^e_i =0, \quad \e_{ijk}\p_j B^e_k = \varepsilon_0 \mu_0 \p_t E^e_i + \mu_0 j^e_i. \label{GED-e}\\
&& \nonumber \\
&&\mbox{\underline{Magnetic limit:}} \nonumber \\
&& \p_i E^m_i = \frac{\rho_m}{\varepsilon_0}, \quad \e_{ijk}\p_j E^m_k = - \p_t B^m_i , \quad \p_i B^m_i =0, \quad \e_{ijk}\p_j B^m_k =  \mu_0 j^m_i. \label{GED-m}
\eea
To connect to our formalism of previous sections, we shall re-derive the above equations \refb{GED-e}, \refb{GED-m} in terms of our scaling limits. We shall work in terms of the gauge fields. We continue to use the non-relativistic scaling of the co-ordinates \refb{contr}. Now we have the additional freedom of choosing the scaling of the gauge field components. We shall scale these in the following way:
\bes \label{lims}
\bea{}
\mbox{Electric limit:} \quad A_0 \to A_0, \, A_i \to \e A_i \label{elimcont}\\
\mbox{Magnetic limit:} \quad A_0 \to \e A_0, \, A_i \to A_i \label{mlimcont}
\eea
\ees
For the moment, let us consider having switched off all sources. It is then easy to check that the source-free Galilean Electrodynamic equations in the two different limits are obtained from the relativistic equations of motion 
$$ \p_\mu F^{\mu \nu} = 0$$
by looking at the above contractions of the gauge fields. Suitably rescaled constants $\varepsilon_0, \mu_0$ give rise to the equations \refb{GED-e} -- \refb{GED-m}. 
The `equations of motion' in the limit defined by \eqref{elimcont} for source-free case are:
\bes{}\label{EOM}
\bea{}
\partial^i \partial_i A_0 &=& 0 \label{eom1}\\
\partial ^j \partial_j A_i - \partial _i \partial_j A^j - \partial_t \partial _i A_0&=& 0 \label{eom2}
\eea
\ees
whereas for the magnetic limit \eqref{mlimcont} these are:
\bes{} \label{magEOM}
\bea{}
\label{mageom1}(\partial ^j \partial_j)A_{i}-\partial_{i}\partial_{j}A^{j} =0\\
\partial_{i}\partial_{t}A^{i}+ (\partial^j \partial_j)A_{0}= 0. \label{mageom2}
\eea
\ees 
We have used the trivial metric $ \delta _{ij}$ on $ \mathbb{R}^3$ for all tensor contractions. Although we need not distinguish between contra or covariant indices, we use them purely for aesthetic reasons.



\subsection{Galilean Conformal Transformations in GED: Contractions} \label{contractionab}

In Sec 2, we looked at the conformal invariance of Maxwell's electromagnetism by referring to the transformations of the fields under the conformal group. The fields $A_\mu$ transformed non-trivially under dilatation and special conformal transformation \refb{transA}. Here we wish to find the non-relativistic versions of these transformations. In Sec 3.3 we worked out, purely from the algebra of Galilean conformal symmetry, its action on $SO(D-1)$ scalars and vectors with definite scaling dimension. However there is another, more straightforward way of looking at this analysis. As pointed out in Sec 3.1, the finite part of GCA can be extracted as a contraction of the relativistic conformal algebra. In Galilean Electrodynamics, we have just discussed two distinct physically meaningful limits the contraction naturally induces on the field space \eqref{lims}. We would exploit these two contractions (space-time ones, reflecting through generators and ones acting on field space) directly on well-known fully relativistic conformal transformation laws to see the parallels of \eqref{sctgen}. More specifically we find the Galilean versions of \eqref{poinc},\eqref{transA} apart from usual Poincare transformations on the relativistic 1-form gauge field in $D$ space-time (Minkowski) dimension. 
As a prescription, Galileanization of transformation rules \eqref{poinc},\eqref{transA} would mean performing the following steps: 
\begin{enumerate}
\item First we perform a canonical space-time split of the transformation formula.
\item Then replace all relevant objects (ie, space and time coordinates, derivative operators with respect to them, field variables) with their transformed version using \eqref{contr} and \eqref{lims}.
\item This would reduce the right hand side of the space-time split version of equations \eqref{poinc}, \eqref{transA} to the form:
$$ \sum _{n= p_1}^{p_2} \epsilon ^n f^{(n)}_{\mathcal{A}} (x, A, \partial A) $$
$p_1 \leq p_2$ and both of them are integers, positive negative or zero, in general. $\mathcal{A}$ stands for arbitrary index structure.
\item In view of the above, now define Galilean version of \eqref{poinc}\eqref{transA} as:
\be{} \tilde{\delta} _\mathcal{A} A  = \lim _{ \e \rightarrow 0 } \e ^{-p_1} \sum _{n= p_1}^{p_2} \epsilon ^n f^{(n)}_\mathcal{A} (x, A, \partial A)
\ee
This last step essentially extracts out the finite part of the relativistic transformation in the limit. 
\end{enumerate}
The two distinct contractions on fields \eqref{lims} are expected to give different transformation rules under the plan of contraction outlined above. Let's tabulate all of them consecutively for the electric limit \eqref{elimcont} first.


\begin{table}[ht]
\vspace{0.3cm}
\centering
\begin{tabular}{|c| l| c|}
\hline
Relativistic  & Galileanized Transformation: Electric version & $p_1$  \\
Transformation &  &  \\
\hline
\hline
	& ${\tilde{\delta}}_0^{\mbox{\tiny{trans}}} A_{0} = \partial _t A_0 =:[H, A_0(t,x)] $ & 0 \\
Translation	& ${\tilde{\delta}}_0^{\mbox{\tiny{trans}}} A_{i} = \partial _t A_i =:[H, A_i(t,x)]$ & 1 \\
			& ${\tilde{\delta}}_i^{\mbox{\tiny{trans}}} A_{0} = -\partial _i A_0 =:[P_i, A_0(t,x)]$ & -1 \\
			& ${\tilde{\delta}}_i^{\mbox{\tiny{trans}}} A_{j} = -\partial _i A_j =:[P_i, A_j(t,x)]$ & 0 \\ 
			\hline
 & ${\tilde{\delta}}_{0 i}^{\mbox{\tiny{Lorentz}}} A_{0} =- t\, \partial _{i}A_{0}=:[B_i, A_0(t,x)] $ & -1\\
Lorentz = \, & ${\tilde{\delta}}_{0 i}^{\mbox{\tiny{Lorentz}}} A_{j} = -t\partial _{i}A_{j} - \delta _{ij} A_0=:[B_i, A_j(t,x)]$  & 0 \\
Boost+ $SO(D-1)$  	& ${\tilde{\delta}}_{ij}^{\mbox{\tiny{Lorentz}}} A_{0} = x_{[i}\partial _{j]}A_{0}=:[J_{ij}, A_0(t,x)] $ & 0 \\
			& ${\tilde{\delta}}_{ij}^{\mbox{\tiny{Lorentz}}} A_{k} = x_{[i}\partial _{j]}A_{k} + \delta _{k[i} A_{j]} =:[J_{ij}, A_k(t,x)]$ & 1 \\
			\hline
Dilatation  	& ${\tilde{\delta}}^{\mbox{\tiny{scale}}} A_{0} = (t \partial _t + x^j\partial_j +\Delta)A_{0} =:[D, A_0(t,x)]$ & 0 \\					
			& ${\tilde{\delta}}^{\mbox{\tiny{scale}}} A_{i} = (t \partial _t + x^j\partial_j+ \Delta)A_{i} =:[D, A_i(t,x)]$ & 1 \\
			\hline
		&${\tilde{\delta}}^{\mbox{\tiny{SCT}}}_0  \, A_{0} = \left[ t^2 \partial_t + 2t\, x^{i} \partial_{i} +2 \Delta t \right]A_{0} =:[K, A_0(t,x)]$& 0\\
Special	&${\tilde{\delta}}^{\mbox{\tiny{SCT}}}_0  \,  A_{i} = \left[ t^2 \partial _{t} + 2t\, x^{j} \partial_{j} +2 \Delta t \right]A_{i}+2 x_i A_0=: [K, A_i(t,x)] $ & 1\\
Conformal            &	 ${\tilde{\delta}}^{\mbox{\tiny{SCT}}}_i A_{0} = - t^2  \partial _i A_0 =: [K_i, A_0(t,x)] $ & -1 \\
			& ${\tilde{\delta}}^{\mbox{\tiny{SCT}}}_i  \, A_{j} = -t^2 \partial _i A_j - 2  \delta_{ij}t A_0 =: [K_i, A_j(t,x)] $ & 0 \\
			\hline
\end{tabular}
\caption{Galilean transformations in the Electric limit}
\label{tableE}
\end{table}

Now let us focus our attention on the two parts of special conformal transformation in its Galileanized version, for the scalar and the vector parts. Comparing with \eqref{sctgen}, it is easy to read out that the arbitrary constants that would label our representations of GCA get indeed fixed as $a_e=-1, b_e=0$ in this particular sector of GED. (Here the subscript on $a,b$ indicates the type of limit in GED.) We also infer that the scaling dimension of the fields $\{A_0, A_i \}$ are given by 
\be{delta}
\Delta = \tilde{\Delta} = \frac{D-2}{2}, 
\ee
i.e. their values are unchanged from the relativistic ones. The set $\{\Delta, a, b\}$ provides the necessary dynamical input required to construct our specific example of the Electric sector of GED. Apart from providing a cross-check of the general analysis we considered earlier in Sec 3.3, this contraction scheme is vitally important to generate the above mentioned dynamical information required to focus attention on the specifics of a particular example of a GCFT.  

The very same analysis is repeated with the magnetic limit \eqref{mlimcont} to extract similar information. The corresponding transformation rules differ from the electric case in some places. This is explicitly displayed in the Magnetic table \ref{tableM} below, which summarises the results of the contraction procedure. 
{\begin{center}
\begin{table}[ht]
\centering
\begin{tabular}{|c| l| c|c|}
\hline
Relativistic  & Galileanized Transformation: Magnetic version & Same as  & $p_1$  \\
Transformation&  & Electric & \\
\hline
\hline
& ${\tilde{\delta}}_0^{\mbox{\tiny{trans}}} A_{0} = \partial _t A_0 =:[H, A_0(t,x)] $ &\checkmark& -1 \\
Translation	& ${\tilde{\delta}}_0^{\mbox{\tiny{trans}}} A_{i} = \partial _t A_i =:[H, A_i(t,x)]$ &\checkmark& 0 \\
			& ${\tilde{\delta}}_i^{\mbox{\tiny{trans}}} A_{0} = -\partial _i A_0 =:[P_i, A_0(t,x)]$ &\checkmark& -2 \\
			& ${\tilde{\delta}}_i^{\mbox{\tiny{trans}}} A_{j} = -\partial _i A_j =:[P_i, A_j(t,x)]$ &\checkmark& -1 \\ 
			\hline
& ${\tilde{\delta}}_{0 i}^{\mbox{\tiny{Lorentz}}} A_{0} =- t\, \partial _{i}A_{0}- A_i =:[B_i, A_0(t,x)] $ &\text{\sffamily X} & 0\\
Lorentz= \,& ${\tilde{\delta}}_{0 i}^{\mbox{\tiny{Lorentz}}} A_{j} = -t\partial _{i}A_{j}=:[B_i, A_j(t,x)]$ &\text{\sffamily X}  & -1 \\
    Boost+ $SO(D-1)$  	& ${\tilde{\delta}}_{ij}^{\mbox{\tiny{Lorentz}}} A_{0} = x_{[i}\partial _{j]}A_{0}=:[J_{ij}, A_0(t,x)] $ &\checkmark& 1 \\
			& ${\tilde{\delta}}_{ij}^{\mbox{\tiny{Lorentz}}} A_{k} = x_{[i}\partial _{j]}A_{k} + \delta _{k[i} A_{j]} =:[J_{ij}, A_k(t,x)]$ &\checkmark& 0 \\
			\hline
Dilatation  	& ${\tilde{\delta}}^{\mbox{\tiny{scale}}} A_{0} = (t \partial _t + x^j\partial_j +\Delta)A_{0} =:[D, A_0(t,x)]$&\checkmark & 1 \\					
			& ${\tilde{\delta}}^{\mbox{\tiny{scale}}} A_{i} = (t \partial _t + x^j\partial_j+ \Delta)A_{i} =:[D, A_i(t,x)]$ &\checkmark& 0 \\
			\hline
		&${\tilde{\delta}}^{\mbox{\tiny{SCT}}}_0  \, A_{0} = \left[ t^2 \partial_t + 2t\, x^{i} \partial_{i} +2 \Delta t \right]            
                         A_{0} + 2 x^i\,A_i=:[K, A_0(t,x)]$ &\text{\sffamily X}& 1\\
Special 	&${\tilde{\delta}}^{\mbox{\tiny{SCT}}}_0  \,  A_{i} = \left[ t^2 \partial _{t} + 2t\, x^{j} \partial_{j} +2 \Delta t \right]A_{i}=: [K, A_i(t,x)]$ &\text{\sffamily X} & 0 \\
 Conformal           &	 ${\tilde{\delta}}^{\mbox{\tiny{SCT}}}_i A_{0} = -t^2  \partial _i A_0 -2 t A_i=: [K_i, A_0(t,x)]$ &\text{\sffamily X} & 0 \\
			& ${\tilde{\delta}}^{\mbox{\tiny{SCT}}}_i  \, A_{j} = -t^2  \partial _i A_j  =: [K_i, A_j(t,x)]$ &\text{\sffamily X} & -1 \\
			\hline
\end{tabular}
\caption{Galilean transformations in the Magnetic limit}
\label{tableM}
\end{table}
\end{center}
}

When compared with \eqref{sctgen}, we gather that $a_m=0, b_m=-1$ are values relevant in the magnetic case. The values of $a$ and $b$ are curiously reversed from the electric case. We will have more to say about this in the next section.

\newpage

\section{Galilean Conformal Symmetries in GED: Intrinsic Analysis}

We have already built up the representation of GCA in its full infinitely extended form in Sec 3, considering specifically spinless (scalar) and spin-1 (actually vector in a fundamental representation on $ \mathfrak{so}(D-1)$) primaries. We observed that these representations are labelled by two additional c-numbers ($a,b$), in addition to scaling dimension and spin of the primaries. The goal of the present section is to study these symmetries in the context of Galilean Electrodynamics discussed above. More specifically, we have two distinct dynamical systems here, namely the electric and the magnetic limits of Galilean Electrodynamics. Before delving into the full infinite GCA we would like to check invariance of the equations of motion describing these distinct systems under the actions of scalar ($K$) and vector parts ($K^i$) of special conformal transformation. This would lead us to two different classes of information. Firstly, as expected, Galilean conformal invariance is observed for space-time dimension $D=4$, as for the relativistic case. Secondly, the two constants, that label our representations also get fixed to real values consistent with those arrived at in Sec~\ref{contractionab} from contraction procedure. After that, invariance under infinite GCA will be presented. Note that by invariance, we would always mean on-shell invariance. We would be dealing with Electric and the Magnetic limit separately in the following analysis for the ease of perusal.

\subsection{Invariance under GCA: Electric limit}\label{checkemlim}
For checking invariance of \eqref{EOM} we would take the following strategy. This concerns checking whether the equations still continue to hold with transformed field variables $\Phi(t,x)$. In other words, if an equation of motion has the schematic form: 
$ \square \Phi(t,x) =J$ then if the following also holds, we would have shown that the equation has the proposed symmetry:
\be{}
\square \delta_{ \bigstar} \Phi(t,x) = \square \lb \bigstar , \Phi (t,x) \rb =0 .
\ee
$\bigstar$ here denotes any transformation generator that is relevant here. In what follows, we would be using \eqref{sctgen} for the expression of $\lb \bigstar , \Phi (t,x) \rb$, keeping in mind the definition: $K=L^{(+1)},  K_i = M_i^{(+1)} $. We note that any source $J$ in the right hand side should anyway be annihilated by transformation operators, since they are non-dynamical. 


\textbf{\underline{Checking for \refb{eom1}:}} It is trivial to see that under scalar and the vector SCT respectively:
\bes\label{checka}
\bea{}
(\partial \cdot \partial) \lb K,A_0 (t,x) \rb &=& 2 b_e \left(2 \partial _i A^i + x^i ( \partial \cdot \partial) A_i \right) \\
(\partial \cdot \partial) \lb K^i ,A_0 (t,x) \rb &=& -2 b_e \,t ( \partial \cdot \partial) A^i  
\eea
\ees
Hence for this equation to be invariant under these transformation, we demand $b_e=0$ as the only possible solution.

\textbf{\underline{Checking for \refb{eom2}}:} 
Checking invariance of this equation under scalar SCT is a somewhat long, but straight-forward exercise. However the vector one is shorter and we display it here:
\bes{}
\bea{}
&&\label{check3} (\partial \cdot \partial) \lb K,A_i \rb - \partial_i  \partial_j \lb K,A^j \rb - \partial_t \partial_ i \lb K, A_0 \rb = -2 (a_e\, D + \Delta -2a_e +1) \partial_i A_0 \non \\ && \hspace{7cm} - 2(a_e+1) x^k \partial_k A_i \\
\label{check4}&& (\partial \cdot \partial) \lb K^i,A_j \rb - \partial_j  \partial_k \lb K^i,A^k \rb - \partial_t \partial_ j \lb K^i, A_0 \rb = 2 (a_e+1) \partial^i \partial_j A_0 \non \\ && \hspace{8cm} - 2b_e \, \partial^i \partial _t( A_j) 
\eea
\ees
Using the solution $b_e=0$ from the invariance condition \eqref{checka} of \eqref{eom1} we infer that $a_e=-1$ should hold for \eqref{check4} to satisfy invariance. Note that these are the exact values of $a_e$ and $b_e$ that we inferred from the results presented in table \ref{tableE}. Substituting this in \eqref{check3} and using \refb{delta}, we get:
\be{} \mbox{Right Hand Side} \, =(D-4) \, \partial _i A_0. \ee
Therefore the equations \eqref{EOM} are Galilean Conformal invariant only at space-time dimension 4.

Having proved Galilean conformal invariance for the finite transformations, it is now straightforward to extend the same analysis for the infinite number of other modes of the GCA. 

\textbf{\underline{Checking for \refb{eom1}:}} It is trivial to see that under $L^{(n)}$ and $M^{(n)}_i$ respectively:
\be{infchecka}
(\partial \cdot \partial) \lb L^{(n)},A_0 (t,x) \rb = 0, \qquad (\partial \cdot \partial) \lb M^{(n)}_i ,A_0 (t,x) \rb = 0  
\ee

\textbf{\underline{Checking for \refb{eom2}}:} A very similar analysis as the previous case now gives
\bes\label{infcheckb}
\bea{}
&&\label{infcheck3}{\hspace{-0.6cm}}(\partial \cdot \partial) \lb L^{(n)},A_i \rb - \partial_i  \partial_j \lb L^{(n)},A^j \rb - \partial_t \partial_ i \lb L^{(n)}, A_0 \rb =- \frac{1}{2}\, n(n+1)\,t^{n-1} (D -4) \partial_i A_0 \\
\label{infcheck4}&&(\partial \cdot \partial) \lb M^{(n)}_i,A_j \rb - \partial_j  \partial_k \lb M^{(n)}_i,A^k \rb - \partial_t \partial_ j \lb M^{(n)}_i, A_0 \rb = 0 
\eea
\ees
Again we conclude that these equations \eqref{EOM} are invariant only at space-time dimension 4 under the modes of the infinitely extended GCA.

\subsection{Invariance under GCA: Magnetic limit}\label{checkmlim}
For GED in magnetic limit, we would proceed in the very same manner as done in the last subsection. It's now obvious that we would be checking how these equations of motion \eqref{magEOM} behave under the scalar and vector special conformal transformation. The results are as follows:

{\textbf{\underline{Checking for \eqref{mageom1}}}}: Under scalar and the vector SCT, we get, respectively:
\bes\label{magchecka}
\bea{}
&&(\partial \cdot \partial) \lb K,A_i (t,x)\rb -(\partial_{j} \partial_{i}) \lb K,A^{j}\rb = -2a_m \lb(D-2) \partial _i + (x \cdot \partial)\partial _i - x_i(\partial \cdot \partial)  \rb A_0 \\
&&(\partial.\partial)[K^{i},A_{j}]-(\partial_{k}\partial_{j})[K^{i},A^{k}] =2a_m t\lb \delta^{i}_{j}(\partial \cdot \partial)-(\partial^{i}\partial_j)\rb A_{0} 
\eea
\ees
Therefore in order to have keep equation invariant under these transformation, we demand $a_m=0$ as the only possible solution.

{\textbf{\underline{Checking for \eqref{mageom2}}}}: Under scalar and vector SCT, the second \eqref{mageom2} magnetic equation yields:
\bes\label{magcheckb}
\bea{}
&&\label{magcheck3}\partial_{t}\partial_{j}[K^{i},A^{j}]+(\partial.\partial)[K^{i},A_{0}] =2a_m(\partial^{i}- t \partial_t \partial^{i} )A_{0}-2(b_m+1)t(\partial \cdot \partial)A^{i}\\
\label{magcheck4}&& \partial_{t}\partial_i [K,A^i]+(\partial \cdot \partial)[K,A_{0}] = (D+4b_m) (\partial \cdot A )+2a_m \left[(D-1)+(x \cdot \partial)\right]\partial_t A_0 \non \\ && \hspace{6cm}+2(b_m+1) x^{j}(\partial \cdot \partial)A_{j}
\eea
\ees
Using the solution $a_m=0$ from the invariance condition \eqref{magchecka} of \eqref{mageom1} we infer that $b_m=-1$ should hold for \eqref{magcheck3} to satisfy invariance. These values are in conformity with those concluded from the analysis of table \ref{tableM}. Substituting this in \eqref{magcheck4}, we get  
\be{} \mbox{Right Hand Side} \,=(D-4) \, \partial \cdot A. \ee
Therefore the magnetic limit equations \eqref{magEOM} too are GC invariant only at space-time dimension $D=4$. 

We see an interesting feature emerging in the above. This is the duality of the parameters labelling GCA representations. The coefficients $a$ and $b$ interchange their values in the electric and magnetic limits. We will have more to say about this shortly. It is now reasonably straightforward to see that the magnetic limit of GED is also invariant under the infinite non-negative modes of GCA:

{\textbf{\underline{Checking for \eqref{mageom1}}}}: The scalar and vector modes of GCA keep this equation invariant only in space-time dimension 4:
\bes{}
\bea{}
&& ( \partial_{j}\partial_{t}[M^{(n)}_{i},A_{j}]+(\partial.\partial)[M^{(n)}_{i},A_{0}])=0\\
&& (\partial_{i}\partial_{t}[L^{(n)},A_{i}]+(\partial.\partial)[L^{(n)},A_{0}])=-\frac{1}{2}n(n+1)(D-4)t^{n-1}(\partial.A) 
\eea
\ees 
{\textbf{\underline{Checking for \eqref{mageom2}}}}: This equation is however invariant under GCA in all dimensions
\bes{}
\bea{}
&&(\partial.\partial)[L^{(n)},A_{i}]-\partial_{i}\partial_{j}[L^{(n)},A_{j}]=0 \\
&&(\partial.\partial)[M^{(n)}_{i},A_{k}]-\partial_{k}\partial_{j}[M^{(n)}_{i},A_{j}]=0
\eea
\ees
The observation we make here is that the magnetic limit, like the electric one is invariant under GCA in 3+1 dimensions only.

The above analysis constitutes the main result of the paper: {\em in both the electric and magnetic limits, the equations of motion of Galilean Electrodynamics have the underlying infinite dimensional symmetry of the Galilean Conformal Algebra \refb{GCA} in $D=4$}. 

Before moving on to an example of the use of this symmetry, a couple of comments are in order. First, let us stress again that this is the first example of a Galilean Conformal Field Theory in $D>2$. 

We have also observed an interesting ``duality" between the electric and magnetic limits. This is the remnant of the parent electro-magnetic duality in the source-free Maxwell's equations. It has been pointed out that Poincare symmetry  is a direct consequence of an electro-magnetic duality \cite{Bunster:2012hm}, and hence our non-relativistic limits individually cannot hope to exhibit this. The exchange of $E_i^e \leftrightarrow B_i^m$ and $B_i^e \leftrightarrow E_i^m$ in the non-relativistic equations reflects the re-emergence of this duality, albeit in a different avatar. This has been reported earlier in \cite{Duval:2014uoa}. In our analysis, we find further evidence of this duality in the exchange of $a_e  \leftrightarrow b_m, b_e  \leftrightarrow a_m$. We shall look at correlation functions below and there as well one would see the exchange of roles of $A_0 \leftrightarrow A_i$ in the two limits indicating the emergence of this duality. 

\subsection{Correlation functions in GED}
Given that the two limits of Galilean Electrodynamics are invariant under the infinite dimensional GCA, we can now call upon the strength of the infinite symmetry to learn more about these systems. Here to give a flavour of what can be calculated, we present the calculation of correlation functions of the field variables $\{ A_0, A_i \}$. We shall restrict ourselves to the finite part of the GCA and look at the two-point functions. It is of interest to look at the full power of the infinite symmetries to write down Ward identities following the case of 2d GCFTs \cite{Bagchi:2009pe}. We defer this to future work. 
In this section we wish to present the correlation functions of our dynamical variables, namely $A_0, A_i$. We start with the assumption that our vacuum state is invariant under the `global' part of GCA, i.e.: $ \{ H, D, P_i, B_i, K, K_i, J_{ij} \}$. Moreover, the assumptions that $A_0$ and $A_i$ are scale $ \Delta$ objects continue here also. We would be presenting the analysis in Electric and Magnetic limit respectively in what follows.

As has been done repeatedly earlier in this article we again take this opportunity to summarize our convention and strategy for calculating the correlators. 
\begin{itemize}
\item[1] Let $ \Phi(t,x)$ and $ \tilde{\Phi}(t,x)$ be two definite spin primaries (in our case spin-0 and 1 objects suffice) with definite scaling dimensions. We would always assume Galilean time-ordering prescription of keeping earlier time operators at the left of any product of operators.
\item[2] Any global generator (assumed to be self-adjoint) $ \bigstar$ being a symmetry of the vacuum, would mean $ \bigstar |0 \> =0$ and $ \< 0 | \bigstar =0  $.
\item[3] When used in the context of correlators, this gives:
$$ \< 0 |\lb \bigstar , \Phi (t_1,x_1) \rb \, \tilde{ \Phi }(t_2, x_2) | 0 \> + \< 0 |  \Phi (t_1,x_1) \, \lb \bigstar , \tilde{ \Phi }(t_2, x_2) \rb | 0 \>=0 $$  
When the transformation rules relevant to our case, as exhibited in Tables \ref{tableE} and  \ref{tableM}, these boil down to differential equations of the correlators. Note that, our transformation rules can mix primary operators in addition to differential action \footnote{For example, in \eqref{dummy} $ \phi_i$ appears in the transformation of $ \phi$ apart from derivatives of $ \phi$}. This results in a set of simultaneous differential equations of correlators. Solutions to these would give us the required correlators.
\end{itemize}

\newpage

\subsubsection*{Correlations in the Electric Limit}

\begin{itemize}
\item $  \underline{ \langle A _0 (t_1, x_1)A _0 (t_2, x_2) \rangle:}$

We want to impose first the invariance under temporal ($H$), spatial ($P_i$) and rotational ($J_{ij}$) transformation of the GCFT vacuum. As expected, it follows from the differential equations thus generated, that:
\be{}
G_{00}(x,t)= \langle A _0 (t_1, x_1)A _0 (t_2, x_2) \rangle = \sum _{m,n \in \mathbb{Z}}c_{m,n} t^m r^n
\ee
where $x^i = x_1^i - x_2^i, \, t =t_1 -t_2 $ and  $ r = x^i x_i $. Now imposing scale invariance, we can restrict it further to (keeping in mind that $A_0$ are scale $ \Delta$ operators):
\be{}
\langle A _0 (t_1, x_1)A _0 (t_2, x_2) \rangle = \sum _{k \in \mathbb{Z}}c_{k} t^{-2 \Delta + k} r^{-k}.
\ee
The next step is to impose the invariance under either of $B_i, K, K_i$. For example, the invariance under $K_i$ results in:
\be{} 
(t^2_1 - t^2_2) \sum _k k\, c_k \, t^{-2 \Delta +k} r^{-k} =0,
\ee
which in turn imply that $c_k =0, ~ \forall \, k\neq 0$. Note that we have used the concerned transformation rules as presented in Table \ref{tableE}, for action of the generators on the dynamical operators. Hence:
\bea{scsc} \langle A _0 (t_1, x_1)A _0 (t_2, x_2) \rangle = c\,t^{-2 \Delta }.\eea
Invariance under rest of the generators ($B_i, K$) give nothing new and simply respect this form of the correlator, as expected.

\item $  \underline{ \langle A _0 (t_1, x_1)A _i (t_2, x_2) \rangle:}$

While using invariance of the vacuum under $H, P_i$ and $J_{ij}$, we keep note of the fact that this correlator must transform as a vector under the action of $J_{ij}$. 
Invariance under $H, P_i, D$ and $J_{ij}$ of the vacuum now dictates, as above
\bea{scvec}G_{0i}(x,t)= \langle A _0 (t_1, x_1)A _i (t_2, x_2) \rangle = \sum _{k}b_{k} \, x^i \, t^{-2 \Delta + k-1} r^{-k}\eea
Let's now implement invariance under $K$ (we could have done with $B_i$ or $K_i$ at this step with same result). We derive:
\bea{scvecK} 
 && \langle 0|\lb K, A _0 (t_1, x_1)A _i (t_2, x_2) \rb |0 \rangle =  \sum _k  b_k t^{-2 \Delta -k-1} [- t ( x^i_1+x^i_2) r^k \non \\ 
 && + k\, r^{k-2} x^i \left(\left(t_1+t_2\right)r^2\right) -2 x_j(t_1 x^j_1 -t_2 x^j_2 )] +2 c\, x^i_2 \, t^{-2 \Delta}
\eea
Here, the coefficient $c$ is same as the one appearing in \eqref{scsc}. For the vanishing of the right hand side of \eqref{scvecK} one may firstly impose $b_k =0, ~ \forall \, k\neq 0$ for simplification. Then the expression becomes:
\be{} 
\lb 2 c \, x^i_2-b_0 (x^i_1+x^i_2) \rb t^{-2 \Delta}.
\ee
As the next step one should demand $c=b_0 =0$ as the only solution for this to be zero \footnote{Note that, for $c=b_0 \neq 0$, we have the expression only functions of $t$ and $x^i$:
$$ \langle \lb K, A _0 (t_1, x_1)A _i (t_2, x_2) \rb \rangle = c x^i t^{-2\Delta} = x^i\,G_{00}(x,t).$$}.
This forces vanishing of both $G_{00} $ and $G_{0i}$. One could have taken any other route, ie, first imposing invariance under $B_i$ or $K_i$. But all of these give the same result.

\item $  \underline{ \langle A _i (t_1, x_1)A _j (t_2, x_2) \rangle:}$

We would see that this is the sole non-zero correlator among the three we are discussing. The most general expression for this (after implementation of $H, P_i, D$ and $J_{ij}$ as symmetry):
\bea{vecvec}
\langle A _i (t_1, x_1)A _j (t_2, x_2) \rangle = \sum_n t^{-2 \Delta -n}r^{n} \lb a_{n} \delta _{ij} + b_n \epsilon_{ijk} \dfrac{x^k}{r}+ c_n \dfrac{x^i x^j}{r^2} \rb \eea
Note that the term $ \epsilon_{ijk} x^k$ is special, occurring only in 3 dimensional case (here our spatial dimension). Had we been dealing with a 3 dimensional relativistic space-time, this term would have to be dropped in favour of principal of micro-causality of Bosonic operators $A_i$.

Now acting this with $K_k$ and using $G_{0i} =0$, we derive that $ b_n = 0= c_n, \, \forall n$ and $a_n = 0,  \, \forall n\neq 0$. This finally yields:
\be{}
\langle A _i (t_1, x_1)A _j (t_2, x_2) \rangle  = a\, \delta _{ij}\, t^{-2 \Delta}.
\ee
\end{itemize}
\subsubsection*{Correlations in the Magnetic Limit}
As expected correlation functions in the Magnetic limit play a role dual to their Electric counterparts. This would be clear as we go along calculating them. As is apparent from transformation properties (Table \ref{tableM}), the vector (spin-1) potential does not involve the scalar one under boost or the scalar and vector conformal transformation. It would therefore be wise to consider the spin-1 spin-1 correlator first.
\begin{itemize}
\item $  \underline{ \langle A _i (t_1, x_1)A _j (t_2, x_2) \rangle:}$

The general form of this correlator, after imposing invariance under $H, P_i, J_{ij}$ and $D$ is same as \eqref{vecvec}. It gets more restricted to the form: 
\bea{mvecvec} 
\langle A _i (t_1, x_1)A _j (t_2, x_2) \rangle = a \, t^{-2 \Delta} \delta_{ij}
\eea 
when one imposes invariance under boost. No more information can be extracted even after checking with $K$ and $ K_i$.

\item $  \underline{ \langle A _0 (t_1, x_1)A _i (t_2, x_2) \rangle:}$

Arguments very similar to those made while deriving \eqref{scvec} can be made here to arrive at (after imposing invariance of $ H, P_i, J_{ij}, D$):
\bea{mscvec}
\langle A _0 (t_1, x_1)A _i (t_2, x_2) \rangle = \sum _{m \in \mathbb{Z}} a_m t^{-2 \Delta - m} \, r^{m-1} \, x_i .
\eea
Invariance under $B_j$ would now mean:
\be{} \sum _{m \in \mathbb{Z}} a_m \, \lb (m-1) (t_1  x^{j}_1 -t_2  x^{j}_2)r^{m-3}\, x_i+ t r^{m-1} \delta_{ij} \rb - a t^{-2 \Delta} \delta_{ij}=0.
\ee
$a$ appearing above is the same one of the right hand side of \eqref{mvecvec}. This is because $B_i$, in the magnetic limit while acting on $A_0$, brings along $A_i$. This is why the spin-1 -- spin-1 correlator becomes relevant in the present analysis. 

Now the above equation is satisfied if $a_m =0 ,\, \forall m \neq 1$ and $a_1 = a$. However, as one wishes to check whether the present form still continues to hold under the invariance condition given by $K_i$, one finds:
\be{} a\, t^{-2 \Delta+1} \delta _{ij} =0.
\ee
Hence we find $a=0$, forcing both the above spin-1 -- spin-1 and scalar -- spin-1 two point functions to vanish.

\item $  \underline{ \langle A _0 (t_1, x_1)A _0 (t_2, x_2) \rangle:}$

Formally, the scalar scalar two point function would involve the above ones in magnetic limit. But as we have just observed, their vanishing nullifies this possibility. Deriving the present correlator would therefore be a repetitive task, already performed for the Electric limit. It is of the form:
\bea{mscsc}
\langle A _0 (t_1, x_1)A _0 (t_2, x_2) \rangle = c\, t^{-2 \Delta}.
\eea
This time, there are no restrictions $c$, though.
\end{itemize}

The above correlators, both in the electric and magnetic case, can be obtained as a limit of the relativistic correlation function \refb{em-corr}. In the electric case, for example, it turns out that some of the correlation functions can have interesting space dependent structures as well. But when we look at the scaling of the gauge field as in \refb{elimcont}, all the correlation functions except $\< A_i A_j \>$ become zero. There is a similar mechanism which matches the correlation functions in the magnetic limit. While we admit that ultra-local correlation functions dependent only on the temporal coordinate is somewhat disappointing, we point out that these are answers dependent on the specific choice of gauge. Gauge transformations remain an intriguing question in the context of non-relativistic electromagnetism. One choice of gauge in the electric limit may be entirely incompatible with the magnetic limit. For example, the Lorentz gauge
$$ \p_\mu A^\mu = 0 $$ 
remains a good choice in the magnetic limit as $x_i \to \e x_i, \, t \to t$ and $A_0 \to \e A_0, \, A_i \to A_i$ keeps this unchanged while this becomes the Coloumb gauge $\p_i A_i = 0$ in the electric limit. We need to have a better understanding gauge invariance in the context of non-relativistic electrodynamics in order to grasp physics behind the form of the correlation functions displayed above. It is possible that we may have imposed stricter conditions than required and missed spatial dependent pieces from our present correlators. We hope to return to the details of this in later work. 


\section{GED in $D=3$}
\subsection{Dual scalar formulation}
It is a curious, but less studied fact that a particular description of Electrodynamics in 3 space-time (Minkowski) dimension also exhibits conformal invariance \cite{Jackiw:2011vz}. This is done by taking the hodge dual of the 3 dimensional Electro-magnetic field strength:
\be{}
\p_{ \mu } \Psi = \dfrac{1}{2} \varepsilon _{ \mu \nu \rho} F^{\nu \rho}.
\ee
The scalar field $\Psi$ defined this way, captures the dynamics of the system. The duality that is inherent to this approach manifests through the reversal of roles of the Bianchi identity and the equation of motion:
\begin{table}[ht]
\centering
\begin{tabular}{|c c c|}
\hline
Equation of motion & & Identity \\
\hline
{} &{}&{} \\ 
$ \p_{\mu}F^{ \mu \nu} =0  $& $ \Leftrightarrow$ & $\varepsilon ^{\mu \nu \rho} \p _{\mu} \p _{\nu} \Psi =0$ \\
{} &{}&{} \\
$ \Box \Psi \equiv \p _{\mu} \p^\mu \Psi =0$ & $ \Leftrightarrow$ & $ \varepsilon ^{ \mu \nu \rho} \p_\mu F_{ \nu \rho} =0$ \\
{} &{}&{} \\
\hline
\end{tabular}
\end{table}

The Lagrangian density in terms of the scalar reads:
\be{}
\mathcal{L} = -\dfrac{1}{4} F^{\mu \nu}F_{\mu \nu} = -\dfrac{1}{2} \p _\mu \Psi
 \p ^\mu \Psi.
 \ee
This is consistent with the equation of motion displayed in the above table. As it represents a theory of a massless scalar, the theory is manifestly scale invariant. Moreover, this fixes the scaling dimension of $ \Psi$ to $1/2$ (by demanding scale invariance of the Lagrangian). It is also easy to see that the theory is conformally invariant under the special conformal transformation considering $ \Psi $ as a dimension $ 1/2$ primary scalar:
\bea{}
\tilde{\delta}_{\mu}^{\mbox{\tiny{SCT}}} \Psi = -x^2 \p_\mu + 2 x_{\mu} x^{\nu} \p_{\nu} + 2 \Delta x_{ \mu} \Psi
\eea
where $ \Delta = \frac{1}{2}$ is the scaling dimension.

The goal of this section is to see whether this invariance continues to hold for Galilean conformal transformations. We would see that the answer is affirmative and it is also invariant under its infinite extension.

In this direction, let us first see the non-relativistic or Galilean version of the equation of motion ($ \Box \Psi =0$ ). We would be using the space-time contraction \eqref{contr} and will not scale the field $ \Psi$. It is easy to see, that any scaling on $ \Psi$ would be redundant as far as the equation of motion is concerned. The equation of motion in Galilean setting reads:
\bea{3deom}
\p_i \p^i \Psi \equiv \nabla^2 \Psi =0.
\eea
\subsection{Invariance under GCA}
Time and space translation generators act on $ \Psi$ in usual manner \eqref{ham}, \eqref{mom}. There is a single component of spatial rotation in this case: $ J:= \frac{1}{2} \varepsilon ^{ij} J_{ij}$. The action of this and the dilatation operator is also easily realized:
\bes \label{d3JD}
\bea{}
\lb J, \Psi (t,x) \rb &=& \varepsilon^{ij} x_i \p _j \Psi (t,x) \\
\lb L^{(0)}, \Psi(t,x)\rb = \lb D, \Psi (t,x) \rb &=& (t \p_t + x^i \p_i + \frac{1}{2}) \Psi (t,x) 
\eea
\ees 
For the action of the boost operator, we take note of the fact that unlike in the $D=4$ case, we do not have any analogue of vector potential in our set of primary fields. More precisely, we can construct our representation of GCA for this particular system, with the single primary field $ \Psi$. This prevents transforming the scalar into a vector, as in \eqref{bphi}. Moreover, invariance under $J$ also prohibits possibility of rapidity labelling the representation. This results in $ \lb M^{(0)}_i, \Psi(0) \rb =0$. Apart from this, $\lb L^{(+1)}, \Psi(0) \rb = 0 = \lb M^{(+1)}_i, \Psi(0) \rb $ are also assumed. Introducing space-time dependence is also fairly straightforward, given that we have already performed the same exercise in Sec \ref{repsec}. The results are:
\bes \label{3dsctgen}
\bea{}
\lb M^{(0)}_i, \Psi (t,x) \rb &=& - t\, \partial_i \Psi (t,x) \\
\lb L^{(+1)}, \Psi(t,x) \rb &=& (t^2 \partial _t + 2 t x^i \partial_i + 2t \Delta) \Psi (t,x) \\
\lb M_i^{(+1)}, \Psi(t,x) \rb &=& - t^2 \, \partial _i  \,\Psi (t,x)
\eea
\ees
From this, we infer that \eqref{3dsctgen} is the special case of \eqref{sctgen} now only with a scalar primary and with $a =0 =b$. We can consider even the infinite extension of these modes, as done earlier in \eqref{inf} for arbitrary spatial dimension. We should keep in mind, while using \eqref{inf}, that the constants $a,b$ appearing there are both zero in the present context. For checking the invariance of the equation of motion \eqref{3deom}, we employ the same tactics as in the Sec \ref{checkemlim} and  \ref{checkmlim}. A number steps involving pure algebraic manipulation with partial derivatives now gives us that:
\bes \label{inf3d}
\bea{}
&&\nabla^2 [L^{(n)}, \Psi] =0 \\
&& \nabla^2 [M_i^{(n)}, \Psi] =0
\eea
\ees
on shell, as before. We therefore conclude that the `hidden' conformal symmetry of 3D Electrodynamics not only holds in its Galilean version, but also it enjoys an infinite dimensional symmetry similar to its counterpart in $D=4$. 


\section{Conclusions and Future Directions}

\subsection{Summary of results}
In this paper, we have initiated a study of symmetries in the non-relativistic limits of Maxwellian electrodynamics and found that infinite dimensional Galilean conformal symmetries emerge in both the Electric and Magnetic limits of Galilean Electrodynamics. 

A central point of our analysis was the construction of a new type of representation theory of the GCA where we considered scale-spin primaries instead of scale-boost primaries considered earlier. We saw how these representations were determined by not only the scaling dimension and the spin of the primaries but also a couple of constants which were fixed by the dynamical information of the particular system under consideration. 

For the specific example of Galilean Electromagnetism, we fixed these constants by implementing a procedure of contraction from the parent relativistic system. With this information, we could examine the symmetries of the equations of motion and found that the Electric and Magnetic limits both exhibited invariance under the infinite dimensional GCA in dimensions $D=4$. Curiously, the values of these constants are the only ones which actually allow the extended symmetries. 

While the invariance under the finite GCA could have been expected from the relativistic Maxwell equations in $D=4$, the invariance under the extended infinite dimensional symmetries is indeed remarkable and we thus found the first example of a (infinitely extended) Galilean Conformal Field Theory in $D>2$. 

We went on to look at correlation functions of the gauge fields to demonstrate in a particular example of how one can exploit the symmetries of Galilean Electrodynamics. Although here we did not use the power of infinite symmetries and restricted our attention to just the finite GCA, this is a problem we wish to return to in the near future. 

We then looked at GED in $D=3$ and its formulation in terms of a scalar field and found that even here Galilean conformal structures emerge as symmetries of the equations of motion. This constitutes another example of a GCFT, now in $D=3$. 

Before we move on, we would like to address a couple of concerns that the reader may have{\footnote{We thank the referee for raising these important questions.}}. 
A potential question could be whether the GCA is the remnant of some infinite dimensional higher spin algebra some free relativistic theories enjoy \cite{sheer}. Let us argue why this is not the case. In two dimensions, as explained in detail in Sec 3.2, the infinite GCA emerges as a contraction of two copies of the Virasoro algebra. The relativistic higher spin algebras in 2d conformal field theories are given by $\mathcal{W}$-algebras. The non-relativistic versions of these algebras are given by In{\"o}n{\"u}-Wigner contractions of these $\mathcal{W}$-algebras. The structures of these non-relativistic $\mathcal{W}$-algebras can e.g. be found in \cite{Afshar:2013vka}. As is evident from this, the GCA forms only the spin-2 version of this algebra and there are infinite families of other generators corresponding to the higher spins. The structure of the contracted higher spin algebras are indeed more rich than just the GCA. In higher dimensions, we expect this feature to hold, i.e. if we were to perform a contraction on a higher spin algebra, the GCA can at most emerge as a sub-algebra of the entire higher spin algebra and there would be additional (infinite number of) generators corresponding to the higher spins. 

This can also be argued from the point of view of non-relativistic holography. The GCA was first discussed in the holographic context in \cite{Bagchi:2009my}, where the dual AdS$_{d+2}$ gravitational theory was shown to be a Newton-Cartan like AdS$_2 \times$ R$^d$ in the non-relativistic limit. The Virasoro sub-algebra of the GCA emerged as the asymptotic symmetries of AdS$_2$. If we suppose that a higher-spin field theory would have a bulk description in terms of a higher spin theory of gravity on AdS in one higher dimension, and we were to look at the non-relativistic limit of this higher spin version of the AdS/CFT correspondence, we would again have a similar AdS$_2 \times$ R$^d$ split of the underlying spacetime. The AdS$_2$ part would now give rise to a part in the higher spin version of the field theory which would now contain a chiral $\mathcal{W}$-algebra \cite{Grumiller:2013swa}. The symmetries of the higher spin non-relativistic field theory would contain the GCA, but along with it an infinite number of other generators corresponding to higher spins. 

We have also exclusively looked at the free Galilean Electromagnetic theory in this paper. The reader may worry whether these extended infinite symmetries are only a feature of the free theory. We are currently studying some interacting theories where GED is coupled with massless matter fields and hope to report on these issues elsewhere. But even in the case where these symmetries belong to just the free theory, we remind the reader of the primary goal of our programme. We wish to uncover a new solvable sector in the AdS$_5$/CFT$_4$ correspondence which would be different from the planar limit of $\mathcal{N}=4$ Super Yang-Mills. If the GCA exists in our generalisations to Yang-Mills and then Super Yang-Mills and is responsible for the integrability of this sector, even if this is restricted to the free theory, this would be a potentially very important result.

\subsection{Connections with Flatspace Holography}

In Sec~\ref{d2}, we had commented on the connection of the GCA to the asymptotic symmetries of flat space. Let us now elaborate on this connection and connected it to our discussions of Galilean Electrodynamics presented in this paper. 
The asymptotic symmetries of flat space at null infinity in three and four dimensions enlarges from the Poincare algebra to what is known as the Bondi-Metzner-Sachs (BMS) algebra. The $\mathfrak{bms}_3$ and $\mathfrak{bms}_4$ algebras are infinite dimensional algebras and the existence of such infinite symmetry structures raises tantalising prospects of applications to holography, which had not been utilised until recently. The $\mathfrak{bms}_3$ is actually isomorphic to the 2d GCA \refb{gca2} with $c_L = 0, c_M = 3/G$ \cite{Bagchi:2010eg} and as mentioned in Sec~\ref{d2}, the mapping between AdS and flatspace is achieved by looking at the Ultra-relativistic contraction \refb{UR}. In two field theory dimensions, the algebra turns out to be isomorphic to the one obtained by non-relativistic contraction \refb{NR} as one can exchange space and time directions without affecting the symmetry structure \cite{Bagchi:2012cy}. This is similar to the isomorphism of the asymptotic symmetry algebras of dS and AdS (of the same dimensions). In bulk dimensions higher than three, this isomorphism does holds in a different sense{\footnote{There is an isomorphism between BMS algebras and what are called ``Semi"- Galilean Conformal Algebras (sGCA) in higher dimensions \cite{Bagchi:2010eg}. These sGCAs are related to the ultra-relativistic ones (CCAs) again by an exchange of spatial and temporal directions.} and one needs to look at Carrolian Conformal Algebras (CCA) as opposed to the GCA.  

In order to link our present explorations of Galilean Electrodynamics to possible dual theories of flat space, it is thus instructive to look at an ultra-relativistic limit of the Maxwellian equations as opposed to the non-relativistic one. In the four dimensional context, we expect that the Carrollian limit will exhibit finite symmetries as the CCA does not have infinite extensions in dimensions higher than three. The presence of similar Electric and Magnetic domains in this ultra-relativistic limit have already been noticed in \cite{Duval:2014uoa}. We expect symmetry enhancements in both these limits. The goal of this particular project is to construct a particular example of flat holography in five bulk dimensions and thereby link it to the AdS$_5$/CFT$_4$ duality by looking at generalisations to $\mathcal{N}=4$ Super Yang Mills and to ultra-relativistic limits in that theory. It is also of interest to explore connections of the results we have established in this work to similar ones in the ultra-relativistic case and see if there exists any potential ``dual" formulation between the non-relativistic and ultra-relativistic sectors similar to the one proposed in \cite{Bagchi:2013bga} in a two-dimensional context for the tensionless limit of string theory. 

The $D=3$ case holds other interesting prospects. Here we can hope for infinite dimensional symmetry enhancements of the scalar formulation of Ultra-Relativistic Electrodynamics to what is the $\mathfrak{bms}_4$ algebra. It would be of interest to explore the extensions to Yang-Mills and Supersymmetric versions if these infinite dimensional actually do exist in the dual scalar formulation in this ultra-relativistic limit. 

Some of the above is work in progress and we hope to report on this soon.

\subsection{Other future directions}

We have already outlined that the next steps in our programme are to first generalise the analysis in this paper to the Yang-Mills theories and then look at supersymmetric versions so as to understand the non-relativistic limit of Super Yang-Mills and study its potential relation to the non-relativistic limit of the AdS$_5$/CFT$_4$ correspondence. We expect the Super-GCA \cite{Bagchi:2009ke, Sakaguchi:2009de, deAzcarraga:2009ch} to play a vital role here. Studies of the SGCA indicate that there are similar infinite dimensional structures that arise in the supersymmetric version. We expect the Galilean versions of Super Yang Mills to exhibit invariance under these extended symmetries.  

Let us now describe some of the other numerous avenues that we wish to explore in relation to this work. 

We have touched upon the debate of scale invariant versus conformally invariant systems in Sec~2 in the relativistic setting and pointed to electromagnetism in $D>4$ as an example of a scale invariant, non-conformal theory. The non-relativistic analogue of this question is whether Galilean invariance coupled with scale invariance in the non-relativistic theory naturally extends to Galilean Conformal invariant systems. We have seen that this does not seem to be the case and Galilean Electrodynamics for $D>4$, like the relativistic case, provides a simple counter-example. It also pointed out earlier in the paper that in two spacetime dimensions, the Zamalodchikov-Polchinski theorem implies that under certain assumptions scale invariance together with Lorentz invariance actually implies conformal invariance. We would like to understand the non-relativistic analogue of this theorem. Among the assumptions of the Zamalodchikov-Polchinski theorem, unitarity plays a vital role. This is a potential stumbling block for us. In $D=2$, using highest weight representations labelled by $L_0$ and $M_0$, we are naturally lead to a non-unitary theory. But there are non-trivial unitary sectors in this theory. It would be good to understand if we are directly lead to such sub-sectors by the non-relativistic Zamalodchikov-Polchinski theorem. 

The Galilean Conformal Algebra is not the only candidate for a non-relativistic conformal algebra. There has been considerable interest in understanding the Schrodinger algebra, the symmetry of the free Schrodinger equations, as a non-relativistic algebra \cite{Hagen:1972pd, Niederer:1972zz, Nishida:2007pj} and this has also been extensively used in the context of non-relativistic holography \cite{Son:2008ye, Balasubramanian:2008dm}. It had been observed in the past that the Schrodinger algebra, like the GCA, can be extended to an infinite dimensional algebra called the Schrodinger-Virasoro algebra \cite{Henkel:1993sg}. Both the GCA and Schrodinger algebras can be cast into the more general structure of $l$-conformal Galilie Algebras \cite{Martelli:2009uc} where $l=1/2$ gives the Schrodinger algebra and $l=1$ leads to the GCA. It is natural to ponder if there could be a more general formulation of Galilean Electrodynamics than what we have presented here. Let us point out here that the Schrodinger algebra cannot be obtained as a contraction of the relativistic conformal algebra and it is much less clear if there can be a formulation of electrodynamics in the non-relativistic limit that is invariant under these symmetries. One possible hurdle is the fact that the Schrodinger algebra contains a mass parameter (a central term in the $[P_i, B_j]$ commutator) and this seems to go against the very nature of electromagnetism. But since we are talking about Galilean electromagnetism, there could be more intricate ways of coming down from the relativistic theory and we defer investigations along this line of thought to future work. 

Non-relativistic physics seems to come hand in hand with a non-metric Newton-Cartan like formulation of spacetime. In the context of the non-relativistic limit of AdS/CFT, it was pointed out in \cite{Bagchi:2009my} that these exotic structures can naturally describe both the bulk and the boundary. In the bulk, the GCA were shown to arise as asymptotic symmetries of this Newton-Cartan structure. This analysis was further elaborated and generalised to include the Schrodinger-Virasoro algebra in \cite{Duval:2009vt}. Of late, there has been a number of works pointing out the importance of Newton-Cartan structures to non-relativistic field theories. See e.g. \cite{Son:2013rqa}. There have also been efforts to understand the importance of these structures to the formulation of non-relativistic supergravity theories \cite{Andringa:2013zja}. It would be instructive to understand the significance of these structures in our formulation of Galilean Electrodynamics and understand how the extended symmetries emerge in this context. The present lack of a Lagrangian description of the two limits could also perhaps be addressed when one looks at these exotic structures and attempts to formulate the problem in a language more intrinsically non-relativistic. 

We have been interested in field theories in $D=4$ in this piece of work. It would also be of interest to look at examples in lower dimensions, e.g. studies of non-relativistic conformal quantum mechanics by studying analogues of the dAFF \cite{de Alfaro:1976je} model in the Galilean context. See \cite{Fedoruk:2011ua} for related work. There have been studies of the holographic duals of the dAFF models in \cite{Chamon:2011xk} as an effort to understand the AdS$_2$/CFT$_1$ connection. It would be very interesting to understand similar holographic structures in the case of the non-relativistic theories as well. What is particularly intriguing about this is that the non-relativistic limit of any AdS$_{d+1}$/CFT$_d$ seems to pick out a particular AdS$_2$/CFT$_1$ connection. So if we were to now concentrate on the limit where $d=1$, then it is far from obvious what one would end up with.

Electromagnetism and Yang-Mills theory, with massless charged particles in $D=4$, has also been in focus recently due to the analysis of asymptotic symmetries \cite{Strominger:2013lka} which turn out to be infinite dimensional at the null boundary of flat space.  It is of interest to understand the question in the non-relativistic limit and see whether this has any link to the infinite structures that we have discussed in this paper. The lack of a Lagrangian description could be a possible stumbling block in attempting the canonical analysis in this case. So the importance of Newton-Cartan structures discussed above again comes to the fore. 

Lastly, we would like to point out a possible relation of our work with the tensionless limit of string theory. In usual tensile string theory, when one fixes conformal gauge in the bosonic closed string theory, there is still some residual gauge freedom which takes the form of conformal invariance on the world sheet. In the tensionless limit of closed bosonic string theory, this residual symmetry turns out to be isomorphic to the 2d GCA and hence it has been suggested that the 2d GCFT would play a central role, very similar to that played by 2d CFTs for tensile strings, in the theory of the tensionless string  \cite{Bagchi:2013bga}. The quantum worldsheet theory of strings leads to the spectrum of physical particles in the spacetime. In particular, the quantising of the open string leads to gauge fields in the spacetime. It is very plausible that the spectrum of the open string in the tensionless limit may lead to the Galilean version of the photon which would be the quantum of Galilean Electromagnetism. In order to venture into this programme, it is obviously important first to construct unambiguously the theory of tensionless open strings in the light of the analysis of the closed string in \cite{Bagchi:2013bga}.

\section*{Acknowledgements}
It is a pleasure to thank Daniel Grumiller and Joan Simon for discussions and comments on the manuscript. We would also like to thank Pulastya Parekh for initial collaboration. AB is partially supported by an INSPIRE award by the Department of Science and Technology, India and a Lise Meitner Award by the FWF Austria. AB would like to thank the string theory groups at University of Santiago de Compostela and ULB Brussels for warm hospitality during the completion of this work. RB thanks IISER Pune for hospitality during a number of visits which resulted in this collaboration.

\end{document}